\pdfoutput=1
\documentclass[aps,prb,twocolumn,superscriptaddress]{revtex4-1}
\usepackage{graphicx}
\usepackage{bm}
\usepackage{amsmath}
\usepackage{amssymb}
\usepackage{euscript}
\usepackage{verbatim}
\usepackage{setspace}
\usepackage{xcolor}
\usepackage{amsfonts}
\usepackage{braket}
\usepackage{subfigure}

\newcommand{\be}{\begin{equation}}
\newcommand{\ee}{\end{equation}}
\newcommand{\bea}{\begin{eqnarray}}
\newcommand{\eea}{\end{eqnarray}}

\allowdisplaybreaks

\newcommand{\Tr}[1]{\mathrm{Tr} #1}

\begin{document}

\title{Multi-impurity chiral Kondo model: correlation functions and anyon fusion rules}
\author{Dor Gabay}
\affiliation{School of Physics and Astronomy, Tel Aviv University, Tel Aviv 6997801, Israel}

\author{Cheolhee Han}
\affiliation{School of Physics and Astronomy, Tel Aviv University, Tel Aviv 6997801, Israel}

\author{Pedro L. S. Lopes}
\affiliation{Department of Physics and Stewart Blusson Institute for Quantum Matter, University of British Columbia, Vancouver, Canada V6T 1Z1}

\author{Ian Affleck}
\affiliation{Department of Physics and Stewart Blusson Institute for Quantum Matter, University of British Columbia, Vancouver, Canada V6T 1Z1}

\author{Eran Sela}
\affiliation{School of Physics and Astronomy, Tel Aviv University, Tel Aviv 6997801, Israel}

\begin{abstract}
	The multichannel Kondo model supports effective anyons on the partially screened impurity, as suggested by its fractional impurity entropy. It was recently demonstrated for the multi-impurity chiral Kondo model, that  scattering of an electron through the impurities depends on the anyon's total fusion channel. Here we study the correlation between impurity-spins. We argue, based on a combination of conformal field theory, a perturbative limit with a large number of channels $k$, and the exactly solvable two-channel case, that the inter-impurity spin correlation  probes the anyon fusion of the pair  of correlated impurities. This may allow, using measurement-only topological quantum computing  protocols, to braid the  multichannel Kondo anyons via consecutive measurements. 
\end{abstract}

\maketitle	

\linespread{1} 

\section{Introduction} 
Multichannel Kondo models display exotic behavior such as anomalous correlation functions and fractional impurity entropy~\cite{affleck1993exact,ludwig1994exact,potok2007observation,andrei1980diagonalization,vigman1980exact,affleck1991universal,han2021entropy}.This residual entropy, in particular, embodies a partial screening of the impurity spin which transforms, otherwise inoffensive, magnetic moments into fractionalized particles; see Fig.~1(a). 

\begin{figure}[h]
	\includegraphics[width=0.5\textwidth]{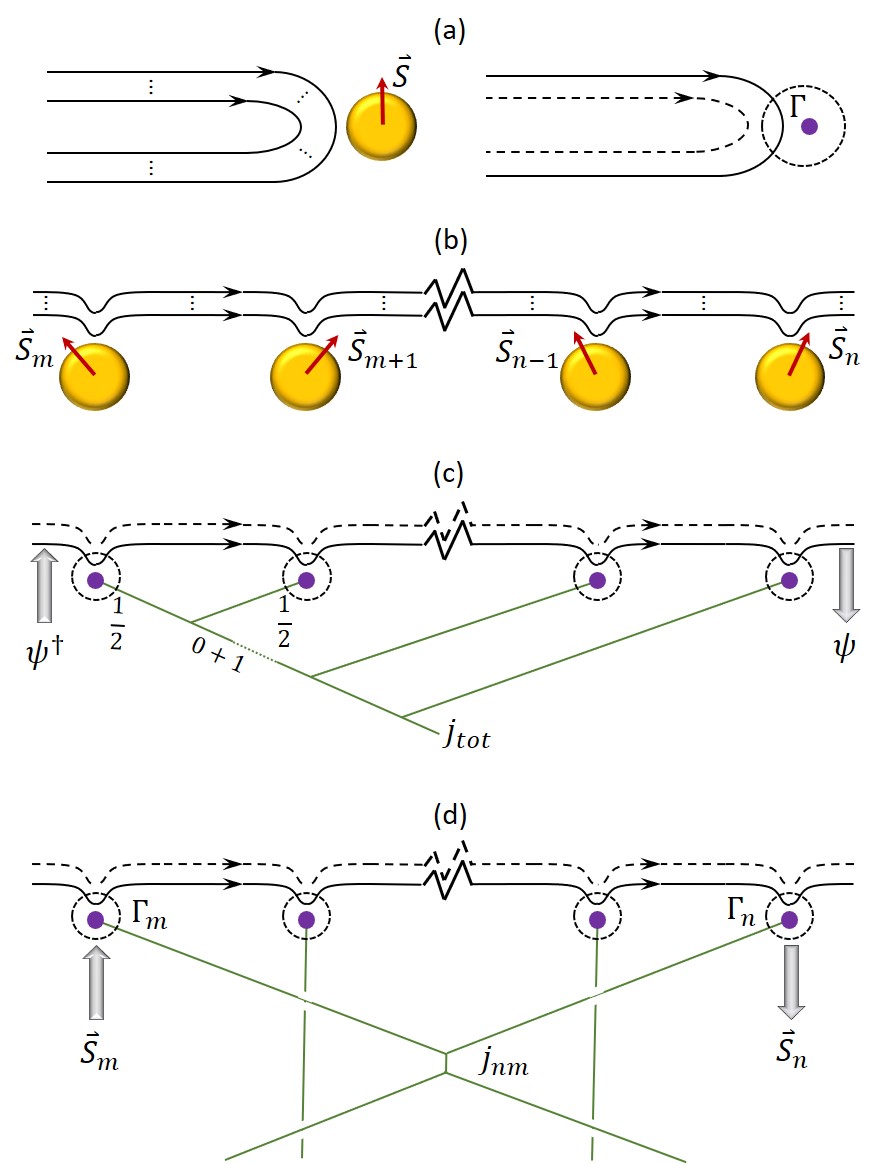}
	\caption{(a) Left: $k-$  channels of spinful conduction electrons interacting with a single impurity spin $\vec{S}$ (yellow filled circle). Right: by virtue of the multi-channel Kondo effect, the impurity spin is partially coupled with the conduction electrons, leaving an anyonic operator denoted $\Gamma$ (blue filled circle). (b) The multi-impurity chiral Kondo model. (c) Correlation functions such as the Green function  $\langle \psi(x_1,t_1) \psi^\dagger(x_2,t_2) \rangle$ depend on the total fusion outcome $j_{tot}$ of the spin-impurity anyons. (d) The impurity-spin correlation function $\langle S_{m}(t_1) S_{n}(t_2) \rangle$, on the other hand, probes the fusion outcome of the corresponding pair of anyons, $j_{nm}$.}
	\label{system}
\end{figure}

Several authors have put forward proposals for using  fractionalized Kondo impurities as building blocks for topological quantum computing~\cite{lopes2020anyons,komijani2020isolating}. Some of us have proposed to do so by leveraging a chiral one-dimensional (1D) multi-impurity Kondo model (c.f. Fig.~1(b)) -- described by $k$ channels of electrons co-propagating through $M$ spin-1/2 impurities, denoted $\vec{S}_n (n=1,\dots,M)$. The key property of the chiral Kondo model is that it allows us to argue that the fractionalized impurity spins behave like non-abelian anyons known from 2D topological phases such as the fractional quantum Hall effect. To make this argument, we  studied the electronic propagator $\langle \psi (x_1,t_1) \psi^\dagger (x_2,t_2) \rangle $. When all the impurities are located at positions $\{x_\ell \}$ between $x_1$ and $x_2$ along the 1D channel, we found that the propagator depended on a quantity shared nonlocally  by the impurities, the total fusion state of $M$ emergent anyons (c.f. Fig. 1(c)). We refer the reader to the literature~\cite{nayak2008non,bonderson2008measurement,kitaev2006anyons,bonderson2012non} for a formal description of the computational space shared by anyons whose basis of states is described by fusion trees, see Fig.~1(c,d).

The development of an architecture for quantum computing involves few, but non-trivial, ingredients: one must identify (i) robust qubits, (ii) a method of quantum control, and (iii) a measurement mechanism for extracting quantum information from the system. Point (i) was solved in our previous work, where a qubit computational space was identified as the nonlocal fusion space of the effective anyons living on the fractionalized impurities. We also proposed an experimental design based on integer quantum Hall systems~\footnote{we take the chance to remark here that Chern insulating systems would, in fact, display less difficulties, with geometrically overlapping channels and no need for magnetic fields to define chiral channels}. To achieve point (ii), we proposed to leverage a measurement-based topological quantum computation~\cite{nayak2008non,bonderson2008measurement,kitaev2006anyons,bonderson2012non} (MBTQC) scheme. The Kondo anyons in our geometry are stuck to impurities, not necessarily easy to braid and maneuver. By opting for a measurement-based scheme, we overcome that difficulty, and reduce the computation process to a problem of state preparation, as long as point (iii) is also solved. The MBTQC approach requires strong measurements. 

Measuring physical quantities, such as the electronic propagator we proposed, provided a way to collapse the anyon state to a given fusion sector. Yet, it was a clumsy solution: for a given choice of $x_1$ and $x_2$, it would only measure the total fusion outcome of all of the multiple anyons between the chosen points of reference.

In this paper we continue the study of the chiral Kondo model and show that measuring certain correlations of impurity spins allows a more direct access to sectors of fixed fusion of desired pairs of fractionalized impurities, hence considerably improving the method for addressing the previously outlined point (iii). Using MBTQC schemes this allows to effectively braid pairs of anyons~\cite{nayak2008non,bonderson2008measurement,kitaev2006anyons,bonderson2012non}.
Our main endeavour is to show that the correlation function of a pair of impurity spins $S_{n}$ and $S_{m}$, filtered to a given SU$(2)$-superselection sector, probes their individual fusion state ($j_{nm}$ in Fig. 1(d)) rather than the total fusion state of the impurities between sites $n$ and $m$ ($j_{tot}$ in Fig. 1(c)). 

As we explain in detail in Sec.~\ref{corrfuncstatedep}, see also  Appendix~\ref{app:correlations}, the correlation functions that we study (as in Ref.~\onlinecite{lopes2020anyons}) are not thermal correlation functions. To appreciated this, note that the total low energy Hilbert space can be separated into the gapless fermionic modes travelling along the chiral mode, and the decoupled fractionalized impurities described by the fusion states. Since the fractionalized impurities are ideally decoupled, they do not have a Hamiltonian and lead to a degeneracy of the ground state which is the computational space. In this space, we do not perform thermal averages but rather assume the state can be prepared, in the sense of (ii), in a given state. For a pair of impurities this state is one of the possible fusion sectors. 
We then study how this correlation function depends on the fusion state.

Since the computation of correlation functions in the Kondo problem is made difficult by the strongly interacting nature of the system, we use complementary methods and exactly solvable limits such as the two channel case~\cite{emery1992mapping}, $k=2$. 
Here we find an opportunity to a technical innovation. Taking note that the electron-impurity coupling constant scales inversely with the number of channels at the non-trivial Kondo fixed points, we obtain another case where exact solution is possible: $k\to \infty$. This way, the large-$k$ limit allows a perturbative treatment of the multi-channel Kondo interactions including the multi-impurity problem. 

The outline of the paper is as follows. In Sec.~\ref{sec:2} we review the key results of Lopes \textit{et. al.}~\cite{lopes2020anyons} and in Sec.~\ref{se:largek} we confirm them in the large-$k$ limit. Building on this approach, in Sec.~\ref{sec:3} we analyze multi-impurity correlations. In Sec.~\ref{sec:4} we  apply refermionization methods for $k=2$, in which the anyons take the form of Majorana fermions. We conclude in Sec.~\ref{sec:5}.

\section{Chiral Kondo model} \label{sec:2}
In this section we  introduce the chiral Kondo model~\cite{lopes2020anyons}, and review the key results on the Green function, the appearance of effective anyons, and the dependence of the Green function on the total anyons fusion state.  More details can be found in Ref.~\onlinecite{lopes2020anyons}.

The chiral $M$-impurity $k$-channel Kondo ($k$CK) system is described by the Hamiltonian $H=H_{0}+H_{K}$. The first term $H_{0}$ is characteristic of $k$ spinful chiral channels
\begin{eqnarray}
H_{0} = -\frac{v_{F}}{2\pi}\int{dx}\psi^{\dagger}i\partial_{x}\psi. \label{eqn:H0}
\end{eqnarray}
Here and below, we consider an infinite chiral system with $\int dx \equiv \int_{-\infty}^\infty dx$. The indices of the spinor $\psi_{i\alpha}$ span $i=1,...,k$ channels (flavors) and spin $\alpha=\uparrow,\downarrow$. The electrons are assumed to be right movers that propagate from infinity and scatter at the positions of the impurities ${x_{1},...,x_{M}}$. The Hamiltonian $H_{K}$ describes the interaction of these spinful channels with $M$ impurity spins $\vec{S}_{\ell}$,
\begin{eqnarray}
H_{K} = \lambda&&\sum_{\ell=1}^{M}\int{dx}\Bigl[\Bigl(\psi^{\dagger}\frac{\bm{\sigma}}{2}\psi\Bigr)\cdot\vec{S}_{\ell}\Bigr](x), \label{eqn:HK} \\
&&\vec{S}_{\ell}(x)=\left( \begin{array}{c} S_{\ell}^{x} \\ S_{\ell}^{y} \\ \Delta S_{\ell}^{z} \end{array} \right)\delta(x-x_{\ell}). \nonumber 
\end{eqnarray}
Here, $\bm{\sigma}$ is the vector of Pauli matrices and $\lambda$ is the exchange coupling constant. The spin-$1/2$ impurities are assumed to be consecutively ordered along the $2k$ chiral spin channels $x_{\ell}<x_{\ell+1}$, as depicted in Fig.~\ref{system}(b), and each contain an anisotropic parameter $\Delta$. Although the spin anisotropy is irrelevant for the low energy behavior~\cite{emery1992mapping}, it will allow for an exact solution of the 2CK systems. 

For a single impurity ($M=1$), Kondo physics is achieved below the Kondo energy $T_{K}=a^{-1} e^{-1/\lambda}$, where $a^{-1}$ is the ultraviolet cutoff. The Kondo energy has a corresponding length scale $\xi_{K}=\hbar v_{F}/k_{B}T_{K}$ dictating the radius at which the spinful electrons are screened along the chiral channels. To alleviate any unwanted correlations in the presence of multiple impurities ($M>1$)~\cite{inPrepLotem}, we will focus on the `dilute' scenario whereby distances between consecutive impurities surpass the Kondo length $|x_{\ell+1}-x_{\ell}|\gg\xi_{K}$. 
Yet, the full linear segment containing all the $M$ impurities, $|x_M-x_1|$ should be smaller than the thermal length $L_T = \hbar v_F/T$ which requires $T < T_K/M$.

\subsection{Boundary condition dependent correlation functions}
Similar to other quantum impurity problems, the chiral Kondo problem can be analysed in the framework of boundary conformal field theory (CFT)~\cite{affleck1993exact,ludwig1994exact}, which allows to compute thermodynamic properties and correlation functions. The free part of the Hamiltonian can be separated into charge, flavour and spin sectors. The Kondo effect takes place in the latter. The spin sector is described by the theory known as SU$(2)_{k}$. Similar to the anyons fusion rules, this theory has primary field denoted as $\Phi_{j}$, with $j=0,1/2,1,...,k/2$. The field $\Phi_{j=1}$ is a vector field for example. The idea is that the Kondo interactions lead to new conformal invariant boundary conditions. They are described by Cardy's boundary CFT~\cite{CARDY_1,CARDY_2,Cardy_Lewellen,Cardy_review}. 

We first review a key result of Affleck and Ludwig for a single impurity at $x=0$.~\cite{ludwig1994exact} Due to the chirality of $H_0$ with a convention of right movers, correlation functions depend only on holomorphic coordinates $z=\tau-ix$. The correlation function of a pair of primary fields $\Phi_j$ spatially located before $\textup{Im}(z_{1})<0$ and after $\textup{Im}(z_{2})>0$ the effective boundary, depend on the (conformal invariant) boundary condition. The boundary conditions themselves are labeled by an index  $i=0,1/2,1,...,k/2$, where $i=0$ corresponds to the trivial boundary condition without the Kondo impurity, and $i=1/2$ is the boudnary condition corresponding to a single Kondo impurity. Other boundary conditions were not realized. Then for conformal invariant boundary condition $i$ the correlation function is~\cite{Cardy_Lewellen} 
\begin{eqnarray}
\braket{\Phi_{j}(z_{1})\Phi_{j}(z_{2})}_{i}=\frac{1}{(z_{1}-z_{2})^{2\Delta_{j}}}\frac{S_{j}^{i}/S_{j}^{0}}{S_{0}^{i}/S_{0}^{0}}. \label{eqn:BCFT}
\end{eqnarray}
Here  $\Delta_{j}=j(j+1)/(2+k)$ is the scaling dimension of the field $\Phi_j$ and the modular-$S$ matrix  is given by~\cite{affleck1993exact} 
\begin{eqnarray}
S_{j}^{j'}=\sqrt{\frac{2}{2+k}}\sin\Biggl[\frac{\pi(2j+1)(2j'+1)}{2+k}\Biggr]. \label{eqn:Smod}
\end{eqnarray}
This result (initially known as the fusion ansatz) of Affleck and Ludwig had been extensively verified in numerous regimes~\cite{affleck1992relevance,pustilnik2004quantum,sela2011exact,mitchell2012universal}, and can be elegantly described by Cardy's boundary CFT~\cite{CARDY_1,CARDY_2,Cardy_Lewellen,Cardy_review} even against experiment~\cite{potok2007observation,keller2015universal}. 

Lopes \textit{et al}.~\cite{lopes2020anyons} generalized the boundary CFT ansatz to multiple impurities. As depicted in Fig.~\ref{system}(c), for multiple spin-$1/2$ impurities, one can achieve a conformal invariant boundary condition $j_{tot} \in \{0, 1/2, 1, ..., k/2 \}$ via multiple fusion
\be
j_{tot} \in    1/2 \times 1/2 \times \dots \times 1/2 .
\ee
We refer to each impurity as a spin-1/2 anyon with fusion rule $1/2 \times 1/2=0+1$ ($k \ge 2$). To treat multiple anyons, one needs to use the  SU$(2)_k$ fusion rules, $j_n \times j_m =|j_n-j_m| + \cdots + \min(j_n+j_m,k-j_n-j_m)$.
Then,
\begin{eqnarray}
\braket{\Phi_{j}(z_{1})\Phi_{j}(z_{2})}_{j_{tot}}=\frac{1}{(z_{1}-z_{2})^{2\Delta_{j}}}\frac{S_{j}^{j_{tot}}/S_{j}^{0}}{S_{0}^{j_{tot}}/S_{0}^{0}}. \label{eqn:corr_asymptotic_M}
\end{eqnarray}
Per the fusion tree of Fig.~\ref{system}(c), the ratio of modular $S$ matrices carries a dependence on the multifusion outcome of an effective anyon $j_{tot}$.

This ansatz was then specified for the case of the fermionic Green function. The fermion field $\psi$ being a spinor, involves the $j=1/2$ spinor field $\Phi_{1/2}$ from SU$(2)_k$. Hence~\cite{lopes2020anyons}
\begin{eqnarray}
\Bigl\langle{\psi_{i\alpha}(z)\psi_{q\beta}^{\dagger}(w)\Bigr\rangle}_{j_{tot}}=\frac{\delta_{i q} \delta_{\alpha \beta}}{z_{1}-z_{2}}\frac{S_{1/2}^{j_{tot}}/S_{1/2}^{0}}{S_{0}^{j_{tot}}/S_{0}^{0}}.
\end{eqnarray}
 For a single channel, $k=1$, the fusion rule $1/2\times1/2=0$ results in a unique fusion outcome. Effectively, all chiral one-channel Kondo models with $M$ odd (even) impurities behave in a similar fashion to one (no) impurity. In these two cases, the ratio of modular $S$-matrices in Eq.~(\ref{eqn:corr_asymptotic_M}), with $j=1/2$ and either $j_{tot}=1/2$ or $0$, give $\mp 1$, where the $-1$ corresponds to the Fermi-liquid $\pi/2$ phase shift of the single channel Kondo effect.

For the two channel case, $k=2$, and an even number of impurities, the ratio of modular $S$ matrices for fermions with $j=1/2$ take two possible values $\pm1$. 
This exemplifies the dependence of an electronic correlation function  on the fusion outcome of the impurities, $1/2\times1/2\times\dots\times1/2=j_{tot}=0,1$ as in Fig.~\ref{system}(c), which acts as two different conformal invariant boundary conditions. 


In the following section, we further demonstrate that the fusion dependence can be reproduced using perturbation theory in the large-$k$ limit.

\section{Large-$k$ Limit}
\label{se:largek}
In this section, we introduce an analytical perturbative approach for the multichannel Kondo model, based on the large $k$-limit. To set the stage for the next sections, we use the large-$k$ approach to show that the results reviewed in the previous section are consistent in this limit.

In the large-$k$ limit, the coupling constant $\lambda$ in the Hamiltonian of Eq.~(\ref{eqn:HK}) becomes gradually smaller, with its renormalization group flow trending as $\lambda=2/k$~\cite{nozieres1980kondo}. The nontrivial outcome of the ratio of $S$-modular matrices can therefore be verified by comparing the lowest-order expansion of $S$-modular matrices to perturbation theory. For a single impurity, this results in a $\mathcal{O}(1/k^{2})$ correction~\cite{affleck1993exact}
\begin{eqnarray}
	\frac{S_{1/2}^{1/2}/S_{1/2}^{0}}{S_{0}^{1/2}/S_{0}^{0}}=1-\frac{3}{2}\frac{\pi^{2}}{k^{2}}+\cdot\cdot\cdot. \label{eqn:fusion_exp}
\end{eqnarray}
A derivation of this result from perturbation theory for fermions, demonstrated in Appendix~\ref{app:asym_singleImp} and performed in Ref.~\onlinecite{affleck1993exact}, serves as a rigorous cross check of the fusion ansatz. 

For multiple impurities, a similar test can be constructed for the multifusion ansatz. Using Eq.~(\ref{eqn:Smod}), the ratio of $S$-modular matrices is expanded for large-$k$ as
\begin{eqnarray}
	\frac{S_{1/2}^{j_{tot}}/S_{1/2}^{0}}{S_{0}^{j_{tot}}/S_{0}^{0}}=1-2\frac{\pi^{2}}{k^{2}}j_{tot}(j_{tot}+1)+\cdot\cdot\cdot. \label{eqn:mutifusion_exp}
\end{eqnarray}
We can now use perturbation theory to demonstrate that, for a two-point fermionic correlator from holomorphic coordinate $\textup{Re}\{w\}$ to $\textup{Re}\{z\}$ with $\textup{Im}\{z\}<0$ and $\textup{Im}\{w\}>0$, the multifusion ansatz correctly generates the $\mathcal{O}(1/k^{2})$ coefficient in Eq.~(\ref{eqn:mutifusion_exp}). To achieve this, we begin by prescribing an ultraviolet cutoff $a$ to the interacting Hamiltonian of Eq.~(\ref{eqn:HK}) 
\begin{eqnarray}
H_{K}=\frac{\lambda}{a^{2}}\sum_{\ell=1}^{M}\int_{-a/2}^{a/2}{dx}\int_{-a/2}^{a/2}{dy}\Bigl[\Bigl(\psi^{\dagger}(x)\frac{\bm{\sigma}}{2}\psi(y)\Bigr)\cdot\vec{S}_{\ell}\Bigr]. \nonumber \\ \label{eqn:HK_ultra}
\end{eqnarray}
Using a path integral approach, we can then expand the exponential of the Kondo action to second-order within the fermionic correlator to arrive at
\begin{widetext}
\begin{eqnarray}
\label{eq:beforepause}\Bigl\langle{\psi_{i\alpha}(z)\psi_{q\beta}^{\dagger}(w)\Bigr\rangle}&&=\Bigl\langle{\psi_{i\alpha}(z)\psi_{q\beta}^{\dagger}(w)\Bigr\rangle}_{(0)}+\frac{\lambda^{2}}{8a^{4}}\sum_{\ell\ell'=1}^{M}\int_{-a/2}^{a/2}{dx_{1}\cdot\cdot\cdot dx_{4}} \label{eqn:asym_largek_0} \\
&&\times\int{d\tau'd\tau''}\Bigl\langle{\psi_{i\alpha}(z)\psi_{q\beta}^{\dagger}(w)\psi_{m\gamma}^{\dagger}(\tau',x_{1})\psi_{m\delta}(\tau',x_{2})\psi_{n\mu}^{\dagger}(\tau'',x_{3})\psi_{n\nu}(\tau'',x_{4})\Bigr\rangle_{(0)}}\sum_{ab}\sigma^{a}_{\gamma\delta}\sigma^{b}_{\mu\nu}\braket{S_{\ell}^{a}(\tau')S_{\ell'}^{b}(\tau'')}. \nonumber 
\end{eqnarray}
\end{widetext}
Here, summation is assumed over repeated indices, and $\langle \dots \rangle_{(i)}$ stands for a  correlation function computed to order $\lambda^i$. Due to the rotational invariance of the impurity spin, its expectation value vanishes (i.e. $\braket{S^{a}}=0$) and the first-order contribution dropped out. 

All the spin-spin correlators computed in this section are of zero order in $\lambda$ and we omit their subscript  $\langle S_{\ell}^{a}(\tau')S_{\ell'}^{b}(\tau'') \rangle_{(0)}$. Yet, before proceeding with the calculation, we make an important  clarification on the nature of these spin correlators.

\subsection{Thermal versus state-dependent 
	\label{corrfuncstatedep}
	spin correlators and expectation values}
In this subsection, before continuing with the evaluation of Eq.~(\ref{eq:beforepause}), we explain that we do not consider $\langle S_{\ell}^{a}(\tau')S_{\ell'}^{b}(\tau'') \rangle$ as thermal correlation functions, but rather, following the ingredients (ii) and (iii) in the introduction, we assume that the system can be prepared in a specific state and ask how that is reflected in the correlations.

To zeroth order in $\lambda$, the spins are decoupled. Then the thermal expectation value for $M$ decoupled spins is defined as
$\braket{A}_{T}=\frac{\Tr{A}}{2^{M}}$.
Consider the correlation function of a pair of decoupled spin-1/2 impurities $\ell \ne \ell'$ appearing in Eq.~(\ref{eqn:asym_largek_3}). 
 Combining their singlet and triplet total angular momentum sectors, their thermal average would naturally result in zero expectation value,
\begin{eqnarray}
&&\Bigl\langle{S_{n}^{a}S_{m}^{b}\Bigr\rangle}_{T}=\frac{\delta^{ab}}{3}\Bigl\langle{\vec{S}_{n}\vec{S}_{m}\Bigr\rangle}_{T} \label{eqn:asym_largek_4} \\
&&=\frac{\delta^{ab}}{6}\Bigl[\Bigl\langle{\bigr(\vec{S}_{n}+\vec{S}_{m}\bigr)^{2}\Bigr\rangle}_{T}-\Bigl\langle{\vec{S}_{n}^{2}\Bigr\rangle}_{T}-\Bigl\langle{\vec{S}_{m}^{2}\Bigr\rangle}_{T}\Bigr] \nonumber \\
&&=\frac{\delta^{ab}}{24}\Tr{\left[\left( \begin{array}{cccc} 0 &  &  &  \\  & 2 &  &  \\  &  & 2 &  \\  &  &  & 2 \end{array} \right)-\frac{3}{2}\mathcal{I}_{4\times 4}\right]}=0,\;(n\ne m). \nonumber
\end{eqnarray}

Instead, we are interested separately in the singlet and triplet states spanned by the two spins in Eq.~(\ref{eqn:asym_largek_4}). To that end, the 2-spin correlator would contain an index $j_{nm}=0,1$ spanning these two states, and
\begin{eqnarray}
\Bigl\langle{S_{n}^{a}(\tau')S_{m}^{b}(\tau'')\Bigr\rangle}_{j_{nm}}=\frac{\delta^{ab}}{6}\Bigl[j_{nm}(j_{nm}+1)-\frac{3}{2}\Bigr]. \label{eqn:asym_largek_5}
\end{eqnarray}
More generally, the expectation value of $M$-impurity spins would take the form
\be
\braket{A}_{j}=\frac{\Tr_{j}{A}}{2j+1},
\ee
where the trace is taken over the $j$-th multiplet of size $(2j+1)$. Thus, below we will use $\langle \dots \rangle_j$ to denote correlation functions in a specific super-selection sector $j$ (whereas we use $\langle \dots \rangle_{(i)}$ to denote different orders in perturbation theory in $\lambda$). 

The total fusion of all the spins, $j_{tot}$, as well as some set of internal fusion states $j_{nm}$, describe the computational state. We ask how correlation functions depend on the computational state.



\subsection{Fusion-dependent Green's function}

We now proceed from Eq.~(\ref{eq:beforepause}) with the calculation of the Green's function.

The zeroth-order contribution is the free-field Green's function
\begin{eqnarray}
\bigl\langle{\psi_{i\alpha}(z)\psi_{q\beta}^{\dagger}(w)\bigr\rangle}_{(0)}=\frac{\delta_{iq}\delta_{\alpha\beta}}{z-w}. \label{eqn:asym_largek_1}
\end{eqnarray}
We can then apply Wick's theorem to the six-point fermionic correlator within the second-order contribution. By keeping only the fully connected contractions, we get 
\begin{widetext}
\begin{eqnarray}
\frac{\lambda^{2}}{8a^{4}}\sum_{\ell\ell'=1}^{M}\int_{-a/2}^{a/2}&&dx_{1}\cdot\cdot\cdot dx_{4}\int{d\tau'd\tau''}\Biggl[\braket{\psi_{i\alpha}(z)\psi_{m\gamma}^{\dagger}(\tau',x_{1})}_{(0)}\braket{\psi_{m\delta}(\tau',x_{2})\psi_{n\mu}^{\dagger}(\tau'',x_{3})}_{(0)}\braket{\psi_{n\nu}(\tau'',x_{4})\psi_{q\beta}^{\dagger}(w)}_{(0)} \label{eqn:asym_largek_2} \\
&&+\braket{\psi_{i\alpha}(z)\psi_{n\mu}^{\dagger}(\tau'',x_{3})}_{(0)}\braket{\psi_{m\delta}(\tau',x_{2})\psi_{q\beta}^{\dagger}(w)}_{(0)}\braket{\psi_{n\nu}(\tau'',x_{4})\psi_{m\gamma}^{\dagger}(\tau',x_{1})}_{(0)}\Biggr]\sum_{ab}\sigma^{a}_{\gamma\delta}\sigma^{b}_{\mu\nu}\braket{S_{\ell}^{a}(\tau')S_{\ell'}^{b}(\tau'')}. \nonumber
\end{eqnarray}
Using the free-field Green's function of Eq.~(\ref{eqn:asym_largek_1}), spin SU(2) symmetry $\braket{S_{\ell}^{a}(\tau')S_{\ell'}^{b}(\tau'')}=\delta^{ab} \braket{S_{\ell}^{z}(\tau')S_{\ell'}^{z}(\tau'')}$, the identity in Eq.~(\ref{eqn:appB_0}), and reversing the integration order $\tau'\leftrightarrow\tau''$ in the second component within the square bracket of Eq.~(\ref{eqn:asym_largek_2}), we arrive at
\begin{eqnarray}
\braket{\psi_{i\alpha}(z)\psi_{q\beta}^{\dagger}(w)}_{(2)}=\frac{\lambda^{2}}{8a^{2}}\sum_{\ell\ell'=1}^{M}\int_{-a/2}^{a/2}dx_{1}dx_{4}\int{d\tau'd\tau''}\frac{6\delta_{iq}\delta_{\alpha\beta}}{(z-\tau')\bigl[\tau'-\tau''-i(x_{1}-x_{4})\bigr](\tau''-w)}\braket{S_{\ell}^{z}(\tau')S_{\ell'}^{z}(\tau'')}. \label{eqn:asym_largek_3}
\end{eqnarray}
\end{widetext}
We ignore any sum of $x_{p}$ (for $p=1,...,4$) with $z,w$ under the assumption that $|\textup{Im}(z)|,|\textup{Im}(w)|\gg a$. We are then left with determining the expectation value of the 2-spin correlator. Before proceeding, we take a short detour clarifying our formal definition of impurity-spin expectation values, see also Appendix~\ref{app:correlations}.

By inserting the two sums within Eq.~(\ref{eqn:asym_largek_3}) into the 2-spin correlator, we recognize that, in an analogous way to Eq.~(\ref{eqn:asym_largek_5}), the total angular momentum sectors of $M$-impurity spins can elegantly be captured 
\begin{eqnarray}
\sum_{\ell\ell'}\Bigl\langle{S_{\ell}^{a}(\tau')S_{\ell'}^{b}(\tau'')\Bigr\rangle}_{j_{tot}}&=&\frac{\delta^{ab}}{3}\Bigl\langle{\Bigl(\sum_{\ell}\vec{S}_{\ell}\Bigr)^{2}\Bigr\rangle}_{j_{tot}} \label{eqn:asym_largek_6} \\
&=&\frac{\delta^{ab}}{3}j_{tot}(j_{tot}+1), \nonumber
\end{eqnarray}
where the index $j_{tot}$ now spans the total angular momentum sectors of $M$-impurities. The spatial coordinates of the impurities within the integral can be assumed to be negligible so long as $\textup{Im}(z),\textup{Im}(w)\gg x_{n}$ for $n=1,...,M$ (see Appendix~\ref{app:asymcorrM_int}). The spatio-temporal integrals of Eq.~(\ref{eqn:asym_largek_3}) can then be evaluated, as is done in Appendix~\ref{app:asymcorrM_int}, to show that the second-order contribution in $1/k$ correctly corresponds to Eq.~(\ref{eqn:mutifusion_exp}),
\begin{eqnarray}
\braket{\psi_{i\alpha}(z)\psi_{q\beta}^{\dagger}(w)}_{j_{tot}}=\frac{\delta_{iq}\delta_{\alpha\beta}}{z-w}\Biggl[1-2\frac{\pi^{2}}{k^{2}}j_{tot}(j_{tot}+1)+\cdot\cdot\cdot\Biggr]. \nonumber \\ \label{eqn:largek_pert}
\end{eqnarray}
The total spin $j_{tot}$ is the large $k$ limit of total anyon fusion channel, following  the $SU(2)_k$ fusion rules. Namely, the large-$k$ approach provides a simple picture for the fusion rules, which become those of conventional $SU(2)$ spins in the large-$k$ limit.

\section{Impurity-Spin Correlations: Large-$k$ limit} \label{sec:3}
In the previous section we focused on the Green's function. This is an example of a correlation function of fields evaluated far from the impurities, which we term ``asymptotic correlators.'' As such, they probe the total fusion state. In this section we  focus, instead, on inter-impurity spin correlations.

\subsection{Statement of main result: inter-impurity correlation and fusion of anyons pairs}
Before diving into the calculations, we outline the main result of this section. 

The picture we demonstrate is that in the chiral multi-impurity Kondo model, the expression $\vec{S} \sim \vec{\Phi}(x=0)$ for a single impurity spin gets generalized to include an operator $\Gamma_{\ell}$ acting on the degenerate Hilbert space spanned by the anyons,
\be
\label{eq:intro}
\vec{S}_{\ell} \sim  \Gamma_{\ell} \vec{\Phi}(x_{\ell}),~~~({\ell}=1,\dots M).
\ee

As the main result, we show that, unlike asymptotic correlations which probe the total fusion sector $j_{tot}$ (Eq.~(\ref{eqn:corr_asymptotic_M})), impurity-spin correlations  probe the fusion outcome of the individual pair of anyons associated with the correlated impurities. We conjecture a general form of the leading order contribution to the impurity-spin correlation function, 
\begin{eqnarray}
\braket{S^{a}_{n}(z_{n})S^{b}_{m}(z_{m})}_{j_{nm}} \approx \delta^{ab}\frac{\mathcal{F}_k(j_{nm})}{(z_{n}-z_{m})^{\frac{4}{2+k}}},\;\;\;(k\ge 2). \nonumber \\ \label{eqn:conjecture}
\end{eqnarray}
Here, $j_{nm}=0,1$ spans the total angular momentum sectors of impurity spins $\vec{S}_{m}$ and $\vec{S}_{n}$. This expression, in accordance with Eq.~(\ref{eq:intro}), shows on the one hand the power law dependence on the coordinates $(z_{n}-z_{m})$ dictated by the primary field $\vec{\Phi}$, and on the other hand carries a dependence on the fusion state. Our large $k$ results below are consistent with
\be
\mathcal{F}_{k \to \infty}(j_{nm}) = \frac{1}{6}j_{nm}(j_{nm}+1)-\frac{1}{4}.
\ee
From this equation, the impurity spin correlator probes the fusion state of the specific pair of anyons, independently of the impurity spins between them. In this section we confirm this conjecture in the large $k$-limit.

\subsection{Single Impurity}
Next we dive into the technical calculations that lead to Eq.~(\ref{eqn:conjecture}). First, we develop the large-$k$ techniques for the single impurity case and only then discuss the multi-impurity case in Sec.~\ref{se:multipleimp}. The impatient reader may skip to the main result in Eq.~(\ref{eqn:Ispin_corrM}).

We begin with the preparatory step of studying the single impurity-spin correlation using perturbation theory in the large-$k$ limit. The intra-impurity-spin correlation can be derived perturbatively to second-order in $\lambda$ by expanding the exponential of the Hamiltonian in path integral form,
\begin{widetext}
\begin{eqnarray}
\braket{S^{a}(\tau_{n})S^{b}(\tau_{m})}=\braket{S^{a}(\tau_{n})S^{b}(\tau_{m})}_{(0)}+\frac{\lambda^{2}}{2}\sum_{cd}\int d\tau'd\tau''\Bigl\langle{S^{a}(\tau_{n})S^{b}(\tau_{m})\Bigl(J^{c}S^{c}\Bigr)(\tau')\Bigl(J^{d}S^{d}\Bigr)(\tau'')\Bigr\rangle}. \nonumber 
\end{eqnarray}
Here, $J^{a}=\frac{1}{2}\psi^{\dagger}\sigma^{a}\psi$ is the spin density field. The zeroth-order contribution is quite trivial, giving 
\begin{eqnarray}
\braket{S^{a}(\tau_{n})S^{b}(\tau_{m})}_{(0)}=\frac{\delta^{ab}}{4}. \label{eqn:Sspin_corr}
\end{eqnarray}
The first-order contribution vanishes by rotational invariance, since $\braket{\vec{J}}=0$. This leaves us with the second order contribution, containing a nontrivial 4-spin correlator
\begin{eqnarray}
\braket{S^{a}(\tau_{n})S^{b}(\tau_{m})}_{(2)}=\frac{k}{4}\lambda^{2}\int d\tau'd\tau''\frac{\delta^{ab}}{(\tau'-\tau''-i\epsilon)^{2}}\sum_{c}\Bigl\langle{S^{a}(\tau_{n})S^{a}(\tau_{m})S^{c}(\tau')S^{c}(\tau'')\Bigr\rangle}, \label{eqn:2order_1}
\end{eqnarray}
\end{widetext}
where we utilize the free spin density field operator product expansion (OPE) 
\begin{eqnarray}
[J^{a}(z),J^{b}(w)]=\frac{k}{2}\frac{\delta^{ab}}{(z-w)^{2}}+\frac{i\epsilon^{abc}J^{c}}{z-w}. \label{eqn:J_OPE} 
\end{eqnarray}
The second term of the OPE does not contribute to Eq.~(\ref{eqn:2order_1}). The infinitesimal value $\epsilon$ is fictitiously prescribed to capture the logarithmic divergence from the integral evaluation done in Appendix~\ref{app:spincorr_int}. It is now a matter of recognizing the various time-orderings resulting from the spin correlator to determine whether a nonzero second-order contribution exists. More generally, the 4-spin correlator can be re-expressed as 
\begin{eqnarray}
\Bigl\langle{S^{a}S^{b}S^{c}S^{d}\Bigr\rangle}=\frac{1}{16}\bigl(\delta^{ab}\delta^{cd}-\delta^{ac}\delta^{bd}+\delta^{ad}\delta^{bc}\bigr). \label{eqn:Sdelta} 
\end{eqnarray}
For $\tau'>\tau''$, the spin indices in Eq.~(\ref{eqn:2order_1}) for the six possible time orderings [numbers $(1)$-$(6)$ depicted in Fig.~\ref{fig:1pair}(a)] can then be expressed as
\begin{eqnarray}
(1)=\sum_{c}\Bigl\langle{S^{a}(\tau_{n})S^{a}(\tau_{m})S^{c}(\tau')S^{c}(\tau'')\Bigr\rangle}&=&\frac{3}{16}, \nonumber \\ 
(2)=\sum_{c}\Bigl\langle{S^{a}(\tau_{n})S^{c}(\tau')S^{a}(\tau_{m})S^{c}(\tau'')\Bigr\rangle}&=&-\frac{1}{16}, \nonumber \\ 
(3)=\sum_{c}\Bigl\langle{S^{a}(\tau_{n})S^{c}(\tau')S^{c}(\tau'')S^{a}(\tau_{m})\Bigr\rangle}&=&\frac{3}{16}, \nonumber \\ 
(4)=\sum_{c}\Bigl\langle{S^{c}(\tau')S^{a}(\tau_{n})S^{c}(\tau'')S^{a}(\tau_{m})\Bigr\rangle}&=&\frac{3}{16}, \nonumber \\ 
(5)=\sum_{c}\Bigl\langle{S^{c}(\tau')S^{a}(\tau_{i})S^{a}(\tau_{m})S^{c}(\tau'')\Bigr\rangle}&=&-\frac{1}{16}, \nonumber \\ 
(6)=\sum_{c}\Bigl\langle{S^{c}(\tau')S^{c}(\tau'')S^{a}(\tau_{n})S^{a}(\tau_{m})\Bigr\rangle}&=&\frac{3}{16}. \label{eqn:2order_2}
\end{eqnarray}
Inserting this result back into Eq.~(\ref{eqn:2order_1}) and evaluating the integral, as is done in Appendix~\ref{app:spincorr_int}, we obtain a logarithmic $k$-dependent correction to the impurity-spin correlator
\begin{eqnarray}
\braket{S^{a}(\tau_{n})S^{b}(\tau_{m})}=\frac{\delta^{ab}}{4}\Biggl[1-\frac{4}{k}\log(\tau_{n}-\tau_{m})+\mathcal{O}\Bigl(\frac{1}{k^{2}}\Bigr)\Biggr]. \nonumber \\ \label{eqn:Ispin_corr}
\end{eqnarray}
Interestingly, using $\vec{S} \sim \mathcal{C} \vec{\Phi}(x=0)$ with some nonuniversal constant $\mathcal{C}$, as in Ref.~\onlinecite{ludwig1994exact}, we have  
\begin{eqnarray}
\langle S^a(\tau_n) S^b(\tau_m) \rangle \propto \delta^{ab}\frac{|\mathcal{C}|^{2}}{(\tau_n - \tau_m)^{\frac{4}{2+k}}}. \label{eqn:Ispin_corr_cft}
\end{eqnarray}
Equation~(\ref{eqn:Ispin_corr}) contains the first two terms in the large-$k$ expansion of this CFT result. In what follows, we show that a similar expression is attained in the multi-impurity scenario, up to a nontrivial factor associated to the total spin sectors of the correlated impurities.

\subsection{Multiple Impurities}
\label{se:multipleimp}
In the multi-impurity scenario, the Kondo Hamiltonian contains a sum over $M$ impurity spins. This sum is carried over to the first and second order perturbation terms. 
\begin{widetext}
As before, the first-order contribution is null due to the expectation value of the spin density. The zeroth- and second-order contributions of the inter-impurity-spin correlator for impurity spins at holomorphic coordinates $z_{n}=\tau_{n}-ix_{n}$ and $z_{m}=\tau_{m}-ix_{m}$ are
\begin{eqnarray}
\braket{S^{a}_{n}(z_{n})S^{b}_{m}(z_{m})}&=&\braket{S^{a}_{n}(\tau_{n})S^{b}_{m}(\tau_{m})}_{(0)} \label{eqn:Ispin_corr_M} \\
&&+\frac{\lambda^{2}}{2}\sum_{\ell\ell'=1}^{M}\sum_{cd}\int d\tau'd\tau''\Bigl\langle{S^{a}_{n}(\tau_{n})S^{b}_{m}(\tau_{m})J^{c}(z')S^{c}_{\ell}(\tau')J^{d}(z'')S^{d}_{\ell'}(\tau'')\Bigr\rangle}. \nonumber 
\end{eqnarray}
The zeroth-order contribution can be evaluated by adopting the result of Eq.~(\ref{eqn:asym_largek_5}), where $j_{nm}=0,1$ represents the different total angular momentum sectors generated by the pair of spins. This procedure is different from tracing over the total angular momentum sectors, as emphasized in the previous section. 

Using the OPE of Eq.~(\ref{eqn:J_OPE}), the second-order contribution to the inter-impurity-spin correlator is
\begin{eqnarray}
\braket{S^{a}_{n}(z_{n})S^{b}_{m}(z_{m})}_{(2)}=\frac{k}{4}\lambda^{2}\sum_{\ell\ell'=1}^{M}\int d\tau'd\tau''\frac{\delta^{ab}}{(\tau'-\tau''-i\Delta{x})^{2}}\sum_{c}\Bigl\langle{S^{a}_{n}(\tau_{n})S^{a}_{m}(\tau_{m})S^{c}_{\ell}(\tau')S^{c}_{\ell'}(\tau'')\Bigr\rangle}, \label{eqn:2orderM_1}
\end{eqnarray}
where $\Delta{x}=x_{\ell}-x_{\ell'}$. 

Evaluating this integral is more cumbersome than the one-impurity case. One has to consider all of the possible impurity configurations within the 4-spin correlator, as is done in Appendix~\ref{app:spincorrM_int}. After evaluating this integral and letting $\lambda$ trend as $2/k$, the zeroth- and second-order contributions to the inter-impurity correlator in the large-$k$ limit are
\begin{eqnarray}
\braket{S^{a}_{n}(\tau_{n})S^{b}_{m}(\tau_{m})}_{j_{nm}}=\delta^{ab}\Bigl(\frac{1}{6}j_{nm}(j_{nm}+1)-\frac{1}{4}\Bigr)\Biggl[1-\frac{4}{k}\log(\tau_{n}-\tau_{m})+\mathcal{O}\Bigl(\frac{1}{k^{2}}\Bigr)\Biggr]. \label{eqn:Ispin_corrM}
\end{eqnarray}
\end{widetext}
This is quite remarkable. We obtain the same expression as the large-$k$ limit of the single impurity scenario, i.e. Eq.~(\ref{eqn:Ispin_corr}), up to a factor $\frac{1}{6}j_{nm}(j_{nm}+1)-\frac{1}{4}$ dictating the total angular momentum sectors of the correlated impurity spins (i.e., either a singlet or triplet corresponding to values of $-1/4$ and $1/12$, respectively). To leading order, Eq.~(\ref{eqn:Ispin_corrM}) can be understood within the framework of CFT by expanding the $\ell$-th impurity-spin $\vec{S}_{\ell}$ as~\cite{ludwig1994exact} 
\begin{eqnarray}
\vec{S}_{\ell}\sim \Gamma_{\ell}\vec{\Phi}. \label{eqn:Ispin_exp} 
\end{eqnarray}
In the large-$k$ limit, inter-impurity-spin correlations, following this expansion, would result in Eq.~(\ref{eqn:conjecture}). The additional dependence on $j_{nm}$ implies that, instead of Eq.~(\ref{eqn:Ispin_corr_cft}), the impurity-spin effectively has an operator $\Gamma_{\ell}$ acting nontrivially on the anyon sector. 

The prescribed impurity-spins allow us to probe their internal fusion-tree states independently of the impurity-spins residing between them. In the following section, this outcome will be demonstrated for two-channel Kondo systems.

\section{Impurity-Spin Correlations - 2CK} \label{sec:4}
In the previous section we made the conjecture Eq.~(\ref{eqn:conjecture}) on the dependence of spin-spin correlations on the fusion state, and showed that it is fully consistent with the perturbative large-$k$ limit. In this sectoin we show that it is also consistent with the exactly solvable $k=2$ case. Below, we dive into explicit calculations. The impatient reader may skip to the main results, Eq.~(\ref{eqn:corr_Ispin_EK_2b}) and (\ref{eqn:corr_Ispin_EK_ii2pm}).

Upon scattering at an impurity, electrons in the $k$CK problem are completely transformed into collective degrees of freedom. Remarkably, using the EK prescription~\cite{emery1992mapping}, which consists of a bosonization and refermionization procedure, this process can be described explicitly. More so, impurity-spin correlations can be calculated exactly for $k=2$. Before calculating these correlation functions, we review the notation introduced in the work of Lopes \textit{et al}.~\cite{lopes2020anyons}. The impatient reader may skip to subsection~\ref{sec:4_corr}. 

We begin by bosonizing~\cite{delft1998bosonize,delft1998kondo}
\begin{eqnarray}
\psi_{i\alpha}(x)&=&a^{-1/2}\kappa_{i\alpha}e^{-i\tilde{\phi}_{i\alpha}(x)}, \label{eqn:boson_identity} 
\end{eqnarray}
where $\kappa_{i\alpha}$ are Klein factors. Channel and spin indices span $i=1,2$ and $\alpha=\uparrow,\downarrow$ , respectively. The spin-channel bosons $\tilde{\phi}_{i\alpha}$ obey the commutation relations
\begin{eqnarray}
[\tilde{\phi}_{i\alpha}(x),\tilde{\phi}_{q\beta}(y)]=i\pi\delta_{iq}\delta_{\alpha\beta}\mathrm{sgn}(x-y), \label{eqn:boson_relations}
\end{eqnarray}
while the Klein factors obey
\begin{eqnarray}
\{\kappa_{i\alpha},\kappa_{q\beta}\}=2\delta_{iq}\delta_{\alpha\beta}. \label{eqn:klein_relations1}
\end{eqnarray}
We can express the Hamiltonians of Eqs.~(\ref{eqn:H0})-(\ref{eqn:HK}) in a more convenient basis by mapping the spin-channel densities to their charge ($c$), spin ($s$), flavor ($f$), and spin-flavor ($sf$) degrees of freedom. The orthogonal transformation 
\begin{eqnarray}
\left( \begin{array}{c} \mathcal{N}_{c} \\ \mathcal{N}_{s} \\ \mathcal{N}_{f} \\ \mathcal{N}_{sf} \end{array} \right)=\frac{1}{2}\left( \begin{array}{cccc} 1 & 1 & 1 & 1 \\ 1 & -1 & 1 & -1 \\ 1 & 1 & -1 & -1 \\ 1 & -1 & -1 & 1 \end{array} \right)\left( \begin{array}{c} \tilde{\mathcal{N}}_{1\uparrow} \\ \tilde{\mathcal{N}}_{1\downarrow} \\ \tilde{\mathcal{N}}_{2\uparrow} \\ \tilde{\mathcal{N}}_{2\downarrow} \end{array} \right) \label{eqn:orthog_trnsfrm}
\end{eqnarray}
is performed, where the spin-channel density $\tilde{\mathcal{N}}_{i\alpha}$ is
\begin{eqnarray}
\tilde{\mathcal{N}}_{i\alpha}=\int\frac{dx}{2\pi}\psi_{i\alpha}^\dagger\psi_{i\alpha}. \label{eqn:density}
\end{eqnarray}
In this alternative basis, the commutation relations for the bosonic fields remain the same
\begin{eqnarray}
[\phi_{\eta}(x),\phi_{\eta'}(y)]=i\pi\delta_{\eta\eta'}\mathrm{sgn}(x-y),\;\;(\eta=c,s,f,sf). \nonumber \\ \label{eqn:boson_relations_eta}
\end{eqnarray}
The charge and flavor degrees of freedom decouple. The new Klein factors satisfy~\cite{delft1998kondo}
\begin{eqnarray}
\kappa_{1\uparrow}^{\dagger}\kappa_{1\downarrow}=\kappa_{sf}^{\dagger}\kappa_{s}^{\dagger}, \label{eqn:klein_tranformation} \\
\kappa_{2\uparrow}^{\dagger}\kappa_{2\downarrow}=\kappa_{sf}\kappa_{s}^{\dagger}, \\
\kappa_{1\uparrow}^{\dagger}\kappa_{2\uparrow}=\kappa_{sf}^{\dagger}\kappa_{f}^{\dagger},
\end{eqnarray}
with anti-commutation relations $\{\kappa_{\eta},\kappa_{\eta'}\}=2\delta_{\eta\eta'}$. 

The strong coupling limit of the bosonized Hamiltonian, not including symmetry breaking perturbations, includes only the spin and spin-flavor degrees of freedom
\begin{eqnarray}
H_{0}=\sum_{\eta}\int\frac{dx}{4\pi}(\partial_{x}\phi_{\eta})^{2}, \label{eqn:H_bosonized_0}
\end{eqnarray}
\begin{eqnarray}
H_{K,+}=&&\frac{\lambda_{+}}{2a}\sum_{\ell=1}^{M}\Bigl(\kappa_{sf}^{\dagger}e^{i\phi_{sf}(x_{\ell})}+\kappa_{sf}e^{-i\phi_{sf}(x_{\ell})}\Bigr) \nonumber \\ 
&&\times\Bigl(S_{\ell}^{-}\kappa_{s}^{\dagger}e^{i\phi_{s}(x_{\ell})}-S_{\ell}^{+}\kappa_{s}e^{-i\phi_{s}(x_{\ell})}\Bigr),  \label{eqn:H_bosonized_Kplus}
\end{eqnarray}
\begin{eqnarray}
H_{K,z}=\lambda_{z}\sum_{\ell=1}^{M}\partial_{x}\phi_{s}(x_{\ell})S_{\ell}^{z}. \label{eqn:H_bosonized_Kz}
\end{eqnarray}
Here, we use $v_{F}=1$. The coupling constants $\lambda_{z}=\Delta{\lambda}$ and $\lambda_{+}=\lambda$ are assumed to be equal for all impurities and we define $S_{\ell}^{\pm}=S_{\ell}^{x}\pm iS_{\ell}^{y}$. 

We can further decouple the spin and spin-flavor degrees of freedom at the Toulouse point $\lambda_{z}=1$ by applying a particular unitary transformation $\mathcal{U}=U_{M}\cdot\cdot\cdot U_{1}$, with $U_{\ell}$ transforming each impurity 
\begin{eqnarray}
U_{\ell}=e^{i\lambda_{z}S_{\ell}^{z}\phi_{s}(x_{\ell})}. \label{eqn:unitary_trnsfrm}
\end{eqnarray}
By carefully considering the chiral bosonic commutation relations, impurity spin commutation relations, and ordering of the impurities, only the spin-flavor remains in the strong-coupling limit of the Hamiltonian. This result can then be refermionized by use of the identity
\begin{eqnarray}
\psi_{\eta}(x)=a^{-1/2}\kappa_{\eta}e^{-i\phi_{\eta}(x)}, \label{eqn:refermionized_identity}
\end{eqnarray}
resulting in a significantly simplified Hamiltonian 
\begin{eqnarray}
H_{0}=-\sum_{\eta}\int\frac{dx}{2\pi}\psi_{\eta}^{\dagger}i\partial_{x}\psi_{\eta}, \label{eqn:H_refermionize_0}
\end{eqnarray}
\begin{eqnarray}
H_{K,+}=\frac{\lambda_{+}}{2\sqrt{a}}\sum_{\ell=1}^{M}\Bigl[\psi_{sf}^{\dagger}(x_{\ell})+\psi_{sf}(x_{\ell})\Bigr](d_{\ell}-d_{\ell}^{\dagger}),  \label{eqn:H_refermionize_Kplus}
\end{eqnarray}
where the complex impurity fermion resembles Jordan-Wigner-like strings
\begin{eqnarray}
d_{\ell}=\kappa_{s}^{\dagger}S_{\ell}^{-}e^{i\pi\sum_{m=\ell+1}^{M}S_{m}^{z}}.  \label{eqn:impurity_fermion}
\end{eqnarray}
Clearly, the Toulouse limit of the 2CK problem in the refermionized form is exactly solvable. In further introducing the Majorana basis 
\begin{eqnarray}
\psi_{\eta}=\frac{\chi_{\eta}^{1}+i\chi_{\eta}^{2}}{\sqrt{2}},\;\;\;d_{\ell}=\frac{a_{\ell}+ib_{\ell}}{\sqrt{2}}, \label{eqn:majorana_basis}
\end{eqnarray}
the Hamiltonian can be further simplified 
\begin{eqnarray}
H=-\int\frac{dx}{4\pi}\chi_{sf}^{1}i\partial_{x}\chi_{sf}^{1}+i\frac{\lambda_{+}}{\sqrt{a}}\sum_{\ell=1}^{M}\chi_{sf}^{1}(x_{\ell})b_{\ell}. \label{eqn:H_majorana}
\end{eqnarray}
In the absence of symmetry breaking perturbations, only the set of Majoranas $b_{1},...,b_{M}$ are strongly coupled, while their fractionalized pairs $a_{1},...,a_{M}$ are entirely free, resulting in a $\frac{1}{2}\textup{log}2$ residual entropy per impurity.

\subsection{Correlations} \label{sec:4_corr}
We now calculate the impurity-spin correlation for the Hamiltonian of Eq.~(\ref{eqn:H_majorana}). We begin with the $z$-component of the impurity spin. It is immediately evident that the unitary transformation has no effect on $S_{\ell}^{z}$. We therefore proceed to expressing the correlator in Majorana form using the identity $S_{\ell}^{z}=ia_{\ell}b_{\ell}$. 
The Majorana $a_{\ell}$ is completely decoupled from the  Hamiltonian  Eq.~(\ref{eqn:H_majorana}), leaving only the strongly coupled Majorana $b_{\ell}$. It has been demonstrated, both for single and multiple impurities, that the latter Majoranas are absorbed by the conduction electrons in the 2CK process.~\cite{sela2009nonequilibrium1,sela2009nonequilibrium2,landau2018charge,lopes2020anyons} This allows us to use the following identity 
\begin{eqnarray}
S^{z}_{\ell} \cong \frac{i}{\pi\lambda_{+}} a_\ell \chi^1_{sf}(z_\ell). \label{eqn:ImpExp_2ck}
\end{eqnarray}
This is an explicit form of Eq.~(1), where the anyonic operator $\Gamma_\ell$ is now represented by a Majorana fermion $a_\ell$. The field $\chi_{sf}^1$ has scaling dimension 1/2 and can be identified with the $z$-component of $\vec{\Phi}$.

By inserting Eq.~(\ref{eqn:ImpExp_2ck}) into the longitudinal inter-impurity-correlation and recognizing that $\chi_{sf}^{1}=(\psi_{sf}^{\dagger}+\psi_{sf})/\sqrt{2}$, we arrive at the correlator
\begin{eqnarray}
\braket{S^{z}_{n}S^{z}_{m}}&=&\frac{\mathcal{P}_{a}}{4i\pi^{2}\lambda_{+}^{2}}\braket{\psi_{sf}(z_{n})\psi_{sf}^{\dagger}(z_{m})} \label{eqn:corr_Ispin_EK_2a} \\
&=&\frac{1}{2i\pi^{2}\lambda_{+}^{2}}\frac{\mathcal{P}_{a}}{z_{n}-z_{m}}. \label{eqn:corr_Ispin_EK_2b}
\end{eqnarray}
Here, $\mathcal{P}_{a}=2ia_{n}a_{m}$ is the parity factor associated to Majoranas $a_{n}$ and $a_{m}$. 
Quite remarkably, independent of the order of the $M$ impurities, the correlation of the two measured impurity spins does not depend on the remaining $M-2$ impurities within the system. 

Notice the factor of $i$ in Eq.~(\ref{eqn:corr_Ispin_EK_2b}) compared to the $k \to \infty$ limit in Eq.~(\ref{eqn:conjecture}). For a single impurity, the longitudinal intra-impurity correlation correctly corresponds to Eq.~(\ref{eqn:Ispin_corr_cft})
\begin{eqnarray}
\braket{S^{z}(\tau_{n})S^{z}(\tau_{m})}=\frac{1}{2\pi^{2}\lambda_{+}^{2}}\frac{1}{\tau_{n}-\tau_{m}}, \label{eqn:corr_Ispin_EK_3} 
\end{eqnarray}
where $a^{2}=\frac{1}{2}$. This does not occur for the inter-impurity correlation of Eq.~(\ref{eqn:corr_Ispin_EK_2b}), where the operator $a_{n}a_{m}=\frac{1}{2i}\mathcal{P}_{a}$ is anti-Hermitian. Such a discrepancy may occur at small-$k$ due to the $k$-dependence of $\mathcal{F}_{k}$ in Eq.~(\ref{eqn:conjecture}). 

The same result can be attained for the transverse impurity-spin correlation $\braket{S_{n}^{+}S_{m}^{-}}$. To exemplify this, we follow the EK prescription, starting by carefully applying the unitary transformation outlined in Eq.~(\ref{eqn:unitary_trnsfrm})
\begin{eqnarray}
\mathcal{U}S_{n}^{\pm}\mathcal{U}^{-1}=S_{n}^{\pm}e^{\pm i\phi_{s}(x_{n})}e^{\mp i\pi\sum_{\ell=n+1}^{M}S_{\ell}^{z}}. \label{eqn:unitary_trnsfrm_S}
\end{eqnarray}
After mapping the impurity spins to their Jordan-Wigner-like strings, the  correlator between impurity spin $S_{n}^{+}$ and $S_{m}^{-}$ at holomorphic coordinates $z_{n}=\tau_{n}-ix_{n}$ and $z_{m}=\tau_{m}-ix_{m}$, respectively, is
\begin{figure}[!ht]
	\raisebox{-\height}{\includegraphics[width=\columnwidth]{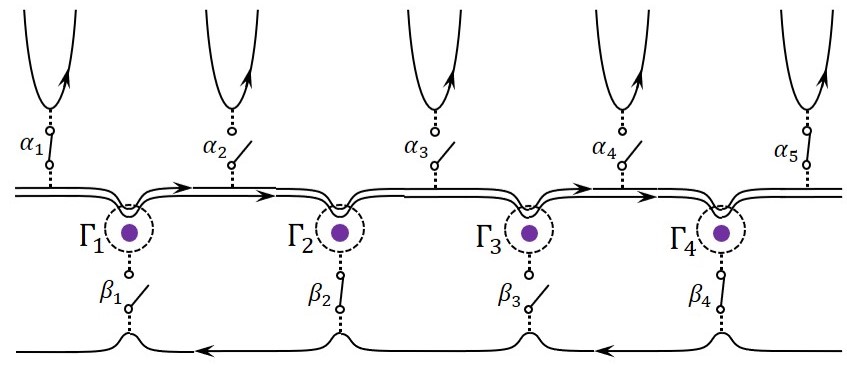}} 
	\caption{Schematic for quantum computing platform depicting $2k$ channels chirally connecting Kondo anyons $\Gamma_{1},...,\Gamma_{4}$. Two methods of probing the system are depicted: (1) at the top, fixed total angular momentum $j_{tot}$ of Kondo anyons residing between weakly coupled contacts is measured through the strongly coupled chiral edge states using switches $\alpha_{1},...,\alpha_{5}$. Here, switches $\alpha_{1}$ and $\alpha_{5}$ are turned on, allowing for a projective measurement of Kondo anyons $\Gamma_{1}\cdots\Gamma_{4}$ in the basis of their total fusion state; (2) At the bottom, fixed total angular momentum sectors $j_{nm}$ of pairs of impurity spins are measured using a weakly coupled chiral edge state controlled by switches $\beta_{1},...,\beta_{4}$. Here, switches $\beta_{2}$ and $\beta_{4}$ apply a projective measurement to Kondo anyons $\Gamma_{2}$ and $\Gamma_{4}$, independent of the Kondo anyons residing between them.}
	\label{fig:1d_model} 
\end{figure}
\begin{eqnarray}
&&\langle{S_{n}^{+}S_{m}^{-}\rangle}= \nonumber \\
&&\braket{S_{n}^{+}e^{i\phi_{s}(z_{n})}e^{-i\pi\sum_{\ell=n+1}^{M}S_{\ell}^{z}}S_{m}^{-}e^{-i\phi_{s}(z_{m})}e^{i\pi\sum_{\ell=m+1}^{M}S_{\ell}^{z}}} \nonumber \\
&&=\braket{d_{n}^{\dagger}d_{m}}\langle{e^{i\phi_{s}(z_{n})}e^{-i\phi_{s}(z_{m})}\rangle}. \label{eqn:corr_Ispin_EK_i2pm}
\end{eqnarray} 
We utilize the identity of Eq.~(\ref{eqn:majorana_basis}) to express the correlator purely in Majorana form and consider only its leading order contribution 
\begin{eqnarray}
\braket{S_{n}^{+}S_{m}^{-}}=\frac{1}{2}\frac{a_{n}a_{m}}{z_{n}-z_{m}}=\frac{1}{4i}\frac{\mathcal{P}_{a}}{z_{n}-z_{m}}, \label{eqn:corr_Ispin_EK_ii2pm}
\end{eqnarray}
where the vertex operator associated to the spin bosonic field gives the conventional CFT result. Clearly, both the longitudinal and transverse impurity-spin correlators result in a parity dependence, analogous to the singlet and triplet total angular momentum sectors in the large-$k$ limit of impurity-spin correlations. 

We now compare the $k=2$ results of Eqs.~(\ref{eqn:corr_Ispin_EK_2b})-(\ref{eqn:corr_Ispin_EK_ii2pm}) with the large-$k$ limit of Eq.~(\ref{eqn:conjecture}). 
On the one hand, $\mathcal{F}_{k\to\infty}(j_{nm})$ in Eq.~(\ref{eqn:conjecture}) takes values $-1/4$ and $1/12$ for fusion channels $j_{nm}=0$ and $1$, respectively. On the other hand, parity eigenvalues for $k=2$ are $\mathcal{P}_a = \pm 1$. This could result from the additional $1/k$ dependence of $\mathcal{F}_{k}(j_{nm})$ emerging at higher-orders in perturbation theory in the large $k-$limit.

\section{Summary} \label{sec:5}
In this paper we studied impurity spin correlation functions of the chiral multi-impurity Kondo model. Substantial evidence has been provided, both from the large-$k$ limit and two-channel case, that, when computed in pure states within specific quantum sectors, impurity spin correlations display a dependence on the internal state nonlocally shared by the effectively fractionalized spins. An interpretation of this dependence has been given in terms of anyon fusion rules. Different than asymptotic correlators like the Green's function that depend only on the total fusion channel~\cite{lopes2020anyons}, 
\begin{figure}[!ht]
	\raisebox{-\height}{\includegraphics[width=\columnwidth]{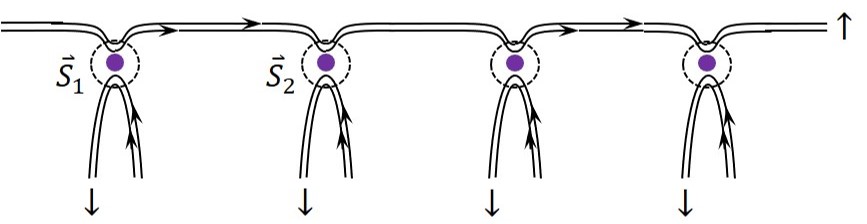}} 
	\caption{A partially connected $k=2$ chiral Kondo system with spin-up ($\uparrow$) channels fully connected and spin-down ($\downarrow$) channels disconnected.}
	\label{fig:system_partial} 
\end{figure}
here the inter-impurity correlations depend on the fusion state of the given pair of correlated impurities.

In contrast to  paradigmatic gapped topological phases as in the fractional quantum Hall effect, our proposed multi-impurity Kondo system is gapless. Nevertheless, the anyons do not couple to unfractionalized degrees of freedom, demonstrating a source of protection. 
To illustrate this, consider for example Majorana fermions in quantum wires~\cite{kitaev2001unpaired,alicea2012new,leijnse2012introduction,beenakker2013search} coupled to gapless modes of a nearby metal. This has been essential for observation of Majorana modes via tunneling,~\cite{mourik2012signatures,das2012zero} as well as in numerous suggestions to probe their properties including braiding~\cite{plugge2016roadmap,vijay2016teleportation,dahan2017non,dahan2020dynamically} and fractional entropy.~\cite{smirnov2015majorana,sela2019detecting} At the same time, the hybridization of Majorana fermions to the surrounding metallic gapless environment leads to decoherence and quasiparticle poissoning.~\cite{rainis2012majorana,albrecht2017transport,karzig2021quasiparticle} In contrast, the two-channel Kondo Majorana fermion is a strongly interacting degree of freedom, and as a result it does not hybridize directly with external Fermi liquid metallic systems. It would be an interesting direction of future inquiry to consider whether the robustness of the isolation of our degrees of freedom survives different types of noise.

Measurement-only schemes~\cite{bonderson2008measurement,bonderson2009measurement,bonderson2012non} require orthogonal projective measurements to be applied onto a sequence of anyons in order to determine the probability of their fusion state. A sequence of such projective measurements generates exchange matrices that can be used for braiding. 
A schematic system allowing to probe different fusion channels is  depicted in Fig.~\ref{fig:1d_model}. Here, switches are used to prescribe a sequence of operations on a set of impurities. At the top, weakly coupled contacts are used to measure a sequence of impurities, while, at the bottom, a specific pair of impurities. For example,  contacts $\alpha_1$ and $\alpha_5$ at the top allow to probe the total fusion of anyons $1,2,3,4$. At the bottom, contacts $\beta_2$ and $\beta_4$ allow to probe the fusion of anyons 2 and 4. 

It should be emphasized that the nontrivial fusion outcome of correlation functions only appears in fully connected system, where all spin-channels propagate in a chiral manner from one impurity to the next, as in Fig.~\ref{system}(b). Fusion of anyonic degrees of freedom does not affect correlation functions within partially connected systems, such as that of Fig.~\ref{fig:system_partial}. This fact can be easily understood from Eq.~(\ref{eqn:conjecture}), where the fusion-dependent factor is multiplied by the correlation of primary fields of different impurities $\braket{\vec{\Phi}^{(n)}(z_{n})\vec{\Phi}^{(m)}(z_{m})}$, which vanishes for dissimilar impurities $n\ne m$ that are partially connected. This requirement for a fully-connected multichannel chiral Kondo model serves as a restriction that should be considered in future experimental setups that could display our predictions.

While here we considered the usual $k$-channel Kondo model with SU$(2)$ symmetry, leading to SU$(2)_k$ anyons, it is interesting to explore anyons in other non-Fermi liquid systems with  emergent symmetries~\cite{ingersent2005kondo,lopez20133,keller2014emergent,mitchell2021so,liberman2021so}.


\section{Acknowledgements} We acknowledge support by European  Research Council (ERC) under the European Unions Horizon 2020 research and innovation programme under grant agreement No. 951541, the US-Israel Binational
Science Foundation (Grant No. 2016255),  ARO (W911NF-20-1-0013), and the Israel Science Foundation grant number 154/19. We thank Heung-Sun Sim for useful discussions.

\appendix

\begin{widetext}
\section{Correlation functions} \label{app:correlations}
In this Appendix, we seek to provide a clear technical explanation on the distinction between thermal correlation functions and the ones we focus on this work. We concentrate on the perturbative large-$k$ case, as a concrete way to state our point. 

We start from the general definition of a thermal correlation function, which can be obtained in imaginary time from
\begin{eqnarray}
\left\langle ...\right\rangle &=&\mathcal{Z}^{-1}\Tr\left[T_{\tau}\left(...\right)e^{-\beta\left(H_{0}+H_{K}\right)}\right],\\
\mathcal{Z}&=&\Tr\left[e^{-\beta\left(H_{0}+H_{K}\right)}\right],
\end{eqnarray}
where $T_{\tau}$ is the time-ordering operator. In the zero-temperature limit, this is nothing but the average of the operators depicted by the ellipsis in the ground state. 
The trace can be computed in a product space between all spin and fermionic degrees of freedom. Applying a path-integral formalism to the fermions, but not the spins, we obtain
\begin{eqnarray}
\left\langle ...\right\rangle &=\frac{\Tr_{\vec{S}}\int\mathcal{D}\left[\psi^{\dagger},\psi\right]\left(...\right)e^{-S\left[\psi^{\dagger},\psi,\left\{ \vec{S}_{\ell}\right\} \right]}}{\Tr_{\vec{S}}\int\mathcal{D}\left[\psi^{\dagger},\psi\right]e^{-S\left[\psi^{\dagger},\psi,\left\{ \vec{S}_{\ell}\right\} \right]}},
\end{eqnarray}
where 
\begin{eqnarray}
S=S_{0}+S_{K}=\int_{-\infty}^{\infty}d\tau\left[\int dx\psi^{\dagger}\partial_{\tau}\psi+H_{0}+H_{K}\right],
\end{eqnarray}
and we took the zero-temperature limit and the remaining trace is only in the impurity degrees of freedom. Perturbation theory follows normally here, resulting in
\begin{eqnarray}
\left\langle ...\right\rangle =\sum_{n=0}^{\infty}\frac{\left(-1\right)^{n}}{n!}\left\langle \left(...\right)S_{K}^{n}\right\rangle _{(0)},
\end{eqnarray}
where
\begin{eqnarray}
\left\langle ...\right\rangle _{(0)}=\frac{\Tr_{\vec{S}}\int\mathcal{D}\left[\psi^{\dagger},\psi\right]\left(...\right)e^{-S_{0}\left[\psi^{\dagger},\psi\right]}}{\Tr_{\vec{S}}\left[1\right]\int\mathcal{D}\left[\psi^{\dagger},\psi\right]e^{-S_{0}\left[\psi^{\dagger},\psi\right]}}.
\end{eqnarray}
For $M$ impurities, the denominator reads $\Tr_{\vec{S}}\left[1\right]=\left(2\times1/2+1\right)^{M}=2^{M}$.

Now, by expanding the order-$n$ term in a multinomial for all the impurities, the arbitrary thermal correlation function reduces to
\begin{eqnarray}
\left\langle ...\right\rangle =\sum_{n=0}^{\infty}\frac{\left(-\lambda\right)^{n}}{n!}\sum_{\left|k\right|=n}\left(\begin{array}{c}
n\\
k
\end{array}\right)\int\left(\prod_{i=0}^{n}d\tau_{i}\right)\mathcal{C}_{\psi}^{k}\left[...\right]\cdot\mathcal{C}_{\vec{S}}^{k}\left[...\right],
\end{eqnarray}
where $k=\left(k_{1},\,...,k_{M}\right)$ and
\begin{eqnarray}
\left(\begin{array}{c}
n\\
k
\end{array}\right)&=&\frac{n!}{k_{1}!\cdots k_{M}!},\;\;\;\left|k\right|\equiv k_{1}+...+k_{M}, \\ 
\left(J^{a}S^{a}\right)^{k}&\equiv&\left(J^{a_{1}}\left(x_{1},\tau_{1}\right)S_{1}^{a_{1}}\left(\tau_{1}\right)\right)^{k_{1}}...\left(J^{a_{M}}\left(x_{M},\tau_{M}\right)S_{M}^{a_{M}}\left(\tau_{M}\right)\right)^{k_{M}} \\ 
&\equiv&\left(J_{1}^{a_{1}}S_{1}^{a_{1}}\right)^{k_{1}}\cdots\left(J_{M}^{a_{M}}S_{M}^{a_{M}}\right)^{k_{M}},
\end{eqnarray}
and $\vec{J}=\frac{1}{2}\psi^{\dagger}\bm{\sigma}\psi$ is the spin current density. Note that a notation is implied here where, whenever the power $k_i=k_1,...,k_M$ is greater than $1$, we also have to introduce different time variables. At perturbation theory of order $n$, $n$ distinct time variables must exist. Finally,
\begin{eqnarray}
\mathcal{C}_{\psi}^{k}\left[...\right]&=\left\langle \left(...\right)_{\psi}\left(J^{a}\right)^{k}\right\rangle _{(0)},\\\mathcal{C}_{\vec{S}}^{k}\left[...\right]&=\left\langle \left(...\right)_{\vec{S}}\left(S^{a}\right)^{k}\right\rangle _{(0)},
\end{eqnarray}
are tensors computed independently for the fermion or spin degrees of freedom.

For spin-spin correlations,
\begin{eqnarray}
\left\langle S_{\ell}^{a}\left(\tau\right)S_{\ell'}^{b}\left(\tau'\right)\right\rangle =\sum_{n=0}^{\infty}\frac{\left(-\lambda\right)^{n}}{n!}\sum_{\left|k\right|=n}\left(\begin{array}{c}
n\\
k
\end{array}\right)\int\left(\prod_{i=0}^{n}d\tau_{i}\right)\mathcal{C}_{\psi}^{k}\left[1\right]\cdot\mathcal{C}_{\vec{S}}^{k}\left[S_{\ell}^{a}\left(\tau\right)S_{\ell'}^{b}\left(\tau'\right)\right].
\end{eqnarray}
For example, the zeroth-order spin piece can be written, in general,
\begin{eqnarray}
\mathcal{C}_{\vec{S}}^{0}\left[S_{\ell}^{a}\left(\tau\right)S_{\ell'}^{b}\left(\tau'\right)\right]=\frac{\delta^{ab}}{6}\sum_{j}\Tr\left[\left(\vec{S}_{\ell}+\vec{S}_{\ell'}\right)^{2}-\frac{3}{2}\mathbb{I}_{2j+1}\right]\begin{cases}
2^{-1} & \ell=\ell',\;\;j=1/2\\
2^{-2} & \ell\neq\ell',\;\;j=0,1
\end{cases}.
\end{eqnarray}
Note that, summing over all $j$ for $\ell\neq\ell'$, $\mathcal{C}_{\vec{S}}^{0}=0$. For a general second order term, we have the spin contribution
\begin{eqnarray}
\mathcal{C}_{\vec{S}}^{2}\left[S_{\ell}^{a}\left(\tau\right)S_{\ell'}^{b}\left(\tau'\right)\right]=2^{-M}\Tr\left[T_{\tau}\left[S_{\ell}^{a}\left(\tau\right)S_{\ell'}^{b}\left(\tau'\right)S_{\mu}^{c}\left(\tau''\right)S_{\nu}^{d}\left(\tau'''\right)\right]\right].
\end{eqnarray}
Although the result of this calculation is basis-independent, if we want to consider the $\ell\neq\ell'$ results, we may pick a basis where we single-out the two external spins $\ell$ and $\ell'$, and consider their sectors of total angular momentum. In this case, we are interested in
\begin{eqnarray}
\mathcal{C}_{\vec{S}}^{2}\left[S_{\ell}^{a}\left(\tau\right)S_{\ell'}^{b}\left(\tau'\right)\right]=\frac{1}{2^{M}}\sum_{j,M_{z},\left\{ s_{p},p\neq \ell,\ell'\right\} }\left\langle \left\{ s_{p}\right\} ,j,M_{z}\left|T_{\tau}\left[S_{\ell}^{a}\left(\tau\right)S_{\ell'}^{b}\left(\tau'\right)S_{\mu}^{c}\left(\tau''\right)S_{\nu}^{d}\left(\tau'''\right)\right]\right|\left\{ s_{p}\right\} ,j,M_{z}\right\rangle _{(0)}. \nonumber \\
\end{eqnarray}
This allows us to get a clear picture. If we want to compare each specific sector at perturbation theory, order by order, we must write
\begin{eqnarray}
\left\langle S_{\ell}^{a}\left(\tau\right)S_{\ell'}^{b}\left(\tau'\right)\right\rangle &&=\sum_{j}\left[\sum_{n=0}^{\infty}\frac{\left(-\lambda\right)^{n}}{n!}\sum_{\left|k\right|=n}\left(\begin{array}{c}
n\\
k
\end{array}\right)\int\left(\prod_{i=0}^{n}d\tau_{i}\right)\mathcal{C}_{\psi}^{k}\left[1\right]\cdot\mathcal{K}_{\vec{S},j}^{k}\left[S_{\ell}^{a}\left(\tau\right)S_{\ell'}^{b}\left(\tau'\right)\right]\right]\\
&&\equiv\sum_{j}\left\langle S_{\ell}^{a}\left(\tau\right)S_{\ell'}^{b}\left(\tau'\right)\right\rangle _{j},
\end{eqnarray}
where
\begin{eqnarray}
\mathcal{K}_{\vec{S},j}^{k}\left[S_{\ell}^{a}\left(\tau\right)S_{\ell'}^{b}\left(\tau'\right)\right]=\frac{1}{2^{M}}\sum_{M_{z},\left\{ s_{p},p\neq \ell,\ell'\right\} }\left\langle \left\{ s_{p}\right\} ,j,M_{z}\left|T_{\tau}\left[S_{\ell}^{a}\left(\tau\right)S_{\ell'}^{b}\left(\tau'\right)\left(\left(S_{1}^{a_{1}}\right)^{k_{1}}...\left(S_{M}^{a_{M}}\right)^{k_{M}}\right)\right]\right|\left\{ s_{p}\right\} ,j,M_{z}\right\rangle _{(0)}. \nonumber \\
\end{eqnarray}
We remark that this decomposition is unique to the computation of a 2-spin correlation function, where $j$ is the total angular momentum of exactly those two spins under study. It provides a link between the thermal correlations and the correlations computed in a given sector $j$ that we consider throughout this work. One can go further: if full quantum control is achieved over the impurity degrees of freedom, the tensor $\mathcal{K}_{\vec{S},j}^{k}$ simplifies to a single piece and must be substituted by
\begin{eqnarray}
\tilde{\mathcal{K}}_{\vec{S},j}^{k}\left[S_{\ell}^{a}\left(\tau\right)S_{\ell'}^{b}\left(\tau'\right)\right]=\frac{1}{2j+1}\sum_{M_{z}}\left\langle \left\{ s_{p}\right\} ,j,M_{z}\left|T_{\tau}\left[S_{\ell}^{a}\left(\tau\right)S_{\ell'}^{b}\left(\tau'\right)\left(\left(S_{1}^{a_{1}}\right)^{k_{1}}...\left(S_{M}^{a_{M}}\right)^{k_{M}}\right)\right]\right|\left\{ s_{p}\right\} ,j,M_{z}\right\rangle _{(0)}. \nonumber \\
\end{eqnarray}
A similar picture holds for the true anyonic case at finite-$k$, by exchanging the traces in the product space of fermions and free spins (which is reasonable in the perturbative case), for a calculation in the interacting ground state where the electronic degrees of freedom decouple from the impurity.

\section{$\braket{\psi_{i\alpha}\psi_{q\beta}^{\dagger}}$ - Single Impurity} \label{app:asym_singleImp}
Given this Appendix is a review of previous results~\cite{affleck1993exact}, we refrain from deriving expressions that are already in the text and instead refer to them. The Green's function for a single impurity, with spatial coordinates $\textup{Im}(w)<0$ and $\textup{Im}(z)>0$, can be calculated to second-order perturbation theory in the large-$k$ limit
\begin{eqnarray}
\Bigl\langle{\psi_{i\alpha}(z)\psi_{q\beta}^{\dagger}(w)\Bigr\rangle}=&&\Bigl\langle{\psi_{i\alpha}(z)\psi_{q\beta}^{\dagger}(w)\Bigr\rangle}_{(0)}+\frac{\lambda}{2}\int{d\tau'}\Bigl\langle{\psi_{i\alpha}(z)\psi_{q\beta}^{\dagger}(w)\psi_{m\gamma}^{\dagger}(\tau')\psi_{m\delta}(\tau')\Bigr\rangle}\sigma^{a}_{\gamma\delta}\braket{S^{a}(\tau')} \label{eqn:appA_0} \\ 
&&+\frac{\lambda^{2}}{8}\int{d\tau'd\tau''}\Bigl\langle{\psi_{i\alpha}(z)\psi_{q\beta}^{\dagger}(w)\psi_{m\gamma}^{\dagger}(\tau')\psi_{m\delta}(\tau')\psi_{n\mu}^{\dagger}(\tau'')\psi_{n\nu}(\tau'')\Bigr\rangle}\sigma^{a}_{\gamma\delta}\sigma^{b}_{\mu\nu}\braket{S^{a}(\tau')S^{b}(\tau'')}. \nonumber 
\end{eqnarray}
The zeroth-order contribution is given by the free field Green's function of Eq.~(\ref{eqn:asym_largek_1}). The first-order contribution is null due to the rotational invariance of the impurity spin $\braket{S^{a}}=0$. After substituting the 2-spin correlator of Eq.~(\ref{eqn:Sspin_corr}), using
\begin{eqnarray}
 \label{eqn:appB_0}
\sum_a \sigma^{a}_{\gamma\delta}\sigma^{a}_{\mu\nu}=2\delta_{\gamma\nu}\delta_{\delta\mu}-\delta_{\gamma\delta}\delta_{\mu\nu}
\end{eqnarray}
and applying Wick's theorem, followed by reordering the time coordinates, we arrive  at the second-order contribution
\begin{eqnarray}
\Bigl\langle{\psi_{i\alpha}(z)\psi_{q\beta}^{\dagger}(w)\Bigr\rangle}_{(2)}=\frac{3\lambda^{2}}{16}\int_{-\infty}^{\infty}{d\tau'd\tau''}\frac{\delta_{iq}\delta_{\alpha\beta}}{(\tau'-\tau'')(\tau'-z)(w-\tau'')}.  
\end{eqnarray}
After some shift of the time coordinates 
\begin{eqnarray}
\Bigl\langle{\psi_{i\alpha}(z)\psi_{q\beta}^{\dagger}(w)\Bigr\rangle}_{(2)}=-\frac{3\lambda^{2}}{16}\int_{-\infty}^{\infty}{dTd\tau}\frac{\delta_{iq}\delta_{\alpha\beta}}{T\bigl(\tau+\frac{1}{2}\Delta\tau-T+ix_{z}\bigr)\bigl(\tau-\frac{1}{2}\Delta\tau+ix_{w}\bigr)}, \label{eqn:appA_1} 
\end{eqnarray}
where $\Delta\tau=\tau_{z}-\tau_{w}$, the integral $\tau$ can be performed to give
\begin{eqnarray}
\int_{-\infty}^{\infty}{d\tau}\frac{\delta_{iq}\delta_{\alpha\beta}}{(\tau+\frac{1}{2}\Delta\tau-T+ix_{z})(\tau-\frac{1}{2}\Delta\tau+ix_{w})}=-\frac{2\pi i}{z-w-T}. 
\end{eqnarray}
Putting this expression back into Eq.~(\ref{eqn:appA_1}) and integrating over $T$, the second-order contribution gives
\begin{eqnarray}
\Bigl\langle{\psi_{i\alpha}(z)\psi_{q\beta}^{\dagger}(w)\Bigr\rangle}_{(2)}&=&-\frac{3\pi i}{8}\lambda^{2}\int_{-\infty}^{\infty}{dT}\frac{\delta_{iq}\delta_{\alpha\beta}}{T(z-w-T)} \nonumber \\
&=&\frac{3\pi i}{16}\lambda^{2}\int_{-\infty}^{\infty}{dT}\frac{\delta_{iq}\delta_{\alpha\beta}}{(T+(z-w)/2)(T-(z-w)/2)}=-\frac{\delta_{iq}\delta_{\alpha\beta}}{z-w}\frac{3\pi^{2}\lambda^{2}}{8}. \label{eqn:appA_2} 
\end{eqnarray}
Setting $\lambda$ to $k/2$ and inserting the zeroth- and second-order results into Eq.~(\ref{eqn:appA_0}), we arrive at a correct correspondence to the fusion ansatz of Eq.~(\ref{eqn:fusion_exp})
\begin{eqnarray}
\Bigl\langle{\psi_{i\alpha}(z)\psi_{q\beta}^{\dagger}(w)\Bigr\rangle}=\frac{\delta_{iq}\delta_{\alpha\beta}}{z-w}\Bigl[1-\frac{3}{2}\frac{\pi^{2}}{k^{2}}+\cdots\Bigr]\approx\frac{\delta_{iq}\delta_{\alpha\beta}}{z-w}\frac{S_{1/2}^{1/2}/S_{1/2}^{0}}{S_{0}^{1/2}/S_{0}^{0}}. \label{eqn:appA_3} 
\end{eqnarray}

\section{$\braket{\psi_{i\alpha}\psi_{q\beta}^{\dagger}}_{(2)}$ - Multiple Impurities}  \label{app:asymcorrM_int}
We demonstrate that the second-order contribution to the asymptotic correlator correctly results in Eq.~(\ref{eqn:largek_pert}). In accounting for the identity Eq.~(\ref{eqn:appB_0}), and the 2-spin correlator of Eq.~(\ref{eqn:asym_largek_6}), the second-order contribution of Eq.~(\ref{eqn:asym_largek_3}) can be written as 
\begin{eqnarray}
\braket{\psi_{i\alpha}(z)\psi_{q\beta}^{\dagger}(w)}_{(2)}&=&\lambda^{2}j_{tot}(j_{tot}+1)\frac{1}{4a^{2}}\int_{-a/2}^{a/2}{dx'dx''}\int_{-\infty}^{\infty}d\tau'\int_{-\infty}^{\infty}d\tau''\frac{\delta_{iq}\delta_{\alpha\beta}}{(z-\tau')\bigl[\tau'-\tau''-i(x'-x'')\bigr](\tau''-w)} \nonumber \\
&=&-\lambda^{2}j_{tot}(j_{tot}+1)\frac{2\pi i}{4a^{2}}\int_{-a/2}^{a/2}{dx'dx''}\theta(x''-x')\int_{-\infty}^{\infty}{d\tau'}\frac{\delta_{iq}\delta_{\alpha\beta}}{(\tau'-w)(\tau'-z)} \nonumber \\
&=&\frac{\lambda^{2}}{4}\frac{\delta_{iq}\delta_{\alpha\beta}}{z-w}j_{tot}(j_{tot}+1)\frac{(2\pi i)^{2}}{a^{2}}\int_{-a/2}^{a/2}dx'\int_{x'}^{a/2}dx''=(2\pi i)^{2}\frac{\lambda^{2}}{8}\frac{\delta_{iq}\delta_{\alpha\beta}}{z-w}j_{tot}(j_{tot}+1). \label{eqn:appB_1}
\end{eqnarray}
In setting $\lambda$ to $2/k$ in the large-$k$ limit, we have
\begin{eqnarray}
\braket{\psi_{i\alpha}(z)\psi_{q\beta}^{\dagger}(w)}_{(2)}=-2\frac{\pi^{2}}{k^{2}}\frac{\delta_{iq}\delta_{\alpha\beta}}{z-w}j_{tot}(j_{tot}+1). \label{eqn:appB_2}
\end{eqnarray}
This concludes our integral evaluation.

\section{$\braket{S^{a}S^{b}}_{(2)}$ - Single Impurity} \label{app:spincorr_int}
We demonstrate that the second-order contribution to the impurity-spin correlator correctly results in Eq.~(\ref{eqn:Ispin_corr}). To simplify the integration process, we recognize that a factor of $3/16$ can be subtracted from all six orderings in Eq.~(\ref{eqn:2order_2}). The subtracted integral, which is constant over all time, clearly vanishes by the residue theorem when substituted into Eq.~(\ref{eqn:2order_1})
\begin{eqnarray}
-\frac{3}{16}\int_{-\infty}^{\infty} d\tau' \int_{-\infty}^{\infty}d\tau'' \frac{1}{(\tau'-\tau''-i\epsilon)^{2}}=0, \label{eqn:2order_3}
\end{eqnarray}
leaving only two regimes of equal contribution with integral expressions of Eq.~(\ref{eqn:2order_2}), each containing a factor of $-1/16-3/16=-1/4$. In further accounting for $\tau'<\tau''$ and exchanging its time coordinates $\tau'\leftrightarrow\tau''$, the integral expression for all contributing regions, i.e. (2) and (5) in Fig.~\ref{fig:1pair}(a) and their mirror images, can be expressed as 
\begin{eqnarray}
I&=&I_{-}+I_{+}=-\frac{1}{2}\int_{\tau_{m}}^{\tau_{n}} d\tau' \int_{\tau_{n}}^{\infty}d\tau'' \frac{1}{(\tau'-\tau''-i\epsilon)^{2}} -\frac{1}{2}\int_{\tau_{m}}^{\tau_{n}} d\tau' \int_{\tau_{n}}^{\infty}d\tau'' \frac{1}{(\tau'-\tau''+i\epsilon)^{2}} 
\end{eqnarray}
Each of the integrals on the right hand side can be evaluated independently
\begin{eqnarray}
I_{\pm}&=&-\frac{1}{2}\Biggl[\int_{-\infty}^{\tau_{n}} d\tau' \int_{\tau_{n}}^{\infty}d\tau'' \frac{1}{(\tau'-\tau''\pm i\epsilon)^{2}}-\int_{-\infty}^{\tau_{m}} d\tau' \int_{\tau_{n}}^{\infty}d\tau'' \frac{1}{(\tau'-\tau''\pm i\epsilon)^{2}}\Biggr] \\
&=&-\frac{1}{2}\Bigl[\log(\tau_{n}-\tau_{m})-\log(\epsilon)\mp\frac{i\pi}{2}\Bigr]. \label{eqn:2order_5}
\end{eqnarray}
The additive constant above cancels out and the divergence can be renormalized, resulting in $I\approx-\log(\tau_{n}-\tau_{m})$. Inserting this result into Eq.~(\ref{eqn:2order_1}) gives the second-order contribution expressed in Eq.~(\ref{eqn:Ispin_corr}). 
\end{widetext}

\section{$\braket{S_{n}^{a}S_{m}^{b}}_{(2)}$ - Multiple Impurities} \label{app:spincorrM_int}
\begin{figure}[!ht]
	\raisebox{-\height}{\includegraphics[width=0.4\textwidth]{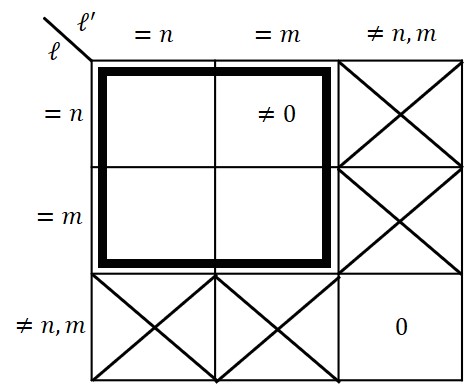}} 
	\caption{All possible impurity-spin configurations of the 4-spin correlator Eq.~(\ref{eqn:2orderM_1}). Configurations that are nonzero, but do not contribute, are marked with an `$\times$'. Configurations that are null are marked with a `$0$'. The boxed set of configurations have a finite contribution.}
	\label{fig:4spin_config} 
\end{figure}

We demonstrate that the second-order contribution to the impurity-spin correlator correctly results in Eq.~(\ref{eqn:Ispin_corrM}). Figure~\ref{fig:4spin_config} compactly summarizes the different possible configuration of the 4-spin correlator in Eq.~(\ref{eqn:2orderM_1}). 

By symmetry of the function we are integrating, it is quite trivial to recognize that the scenario $n\ne\ell\ne\ell'\ne m$ does not differentiate between the different time orderings and is hence zero. 

The next of these scenarios is one of the four off diagonal terms in the last row and last column, marked by an $\times$ in Fig.~\ref{fig:4spin_config}, in which only a single pair of impurities is identical. Their individual contributions are not zero, but the sum of their contributions is indeed zero. To see this, we exemplify the different time orderings numbered $(1)$-$(6)$ in Fig.~\ref{fig:1pair}(a) for the case \{$n \ne m$; $\ell=n$; $\ell'\ne m,n$\}, 
\begin{eqnarray}
(1)=\Bigl\langle{S^{a}_{n}S^{a}_{m}S^{c}_{n}S^{c}_{\ell'}\Bigr\rangle}&=&\frac{1}{4}\delta^{ac}\Bigl\langle{S^{a}_{m}S^{c}_{\ell'}\Bigr\rangle}-\frac{i}{2}\epsilon_{cab}\Bigl\langle{S^{b}_{n}S^{a}_{m}S^{c}_{\ell'}\Bigr\rangle}, \nonumber \\ 
(2)=\Bigl\langle{S^{a}_{n}S^{c}_{n}S^{a}_{m}S^{c}_{\ell'}\Bigr\rangle}&=&\frac{1}{4}\delta^{ac}\Bigl\langle{S^{a}_{m}S^{c}_{\ell'}\Bigr\rangle}-\frac{i}{2}\epsilon_{cab}\Bigl\langle{S^{b}_{n}S^{a}_{m}S^{c}_{\ell'}\Bigr\rangle}, \nonumber \\ 
(3)=\Bigl\langle{S^{a}_{n}S^{c}_{n}S^{c}_{\ell'}S^{a}_{m}\Bigr\rangle}&=&\frac{1}{4}\delta^{ac}\Bigl\langle{S^{a}_{m}S^{c}_{\ell'}\Bigr\rangle}-\frac{i}{2}\epsilon_{cab}\Bigl\langle{S^{b}_{n}S^{a}_{m}S^{c}_{\ell'}\Bigr\rangle}, \nonumber \\ 
(4)=\Bigl\langle{S^{c}_{n}S^{a}_{n}S^{a}_{m}S^{c}_{\ell'}\Bigr\rangle}&=&\frac{1}{4}\delta^{ac}\Bigl\langle{S^{a}_{m}S^{c}_{\ell'}\Bigr\rangle}+\frac{i}{2}\epsilon_{cab}\Bigl\langle{S^{b}_{n}S^{a}_{m}S^{c}_{\ell'}\Bigr\rangle}, \nonumber \\ 
(5)=\Bigl\langle{S^{c}_{n}S^{a}_{n}S^{c}_{\ell'}S^{a}_{m}\Bigr\rangle}&=&\frac{1}{4}\delta^{ac}\Bigl\langle{S^{a}_{m}S^{c}_{\ell'}\Bigr\rangle}+\frac{i}{2}\epsilon_{cab}\Bigl\langle{S^{b}_{n}S^{a}_{m}S^{c}_{\ell'}\Bigr\rangle}, \nonumber \\ 
(6)=\Bigl\langle{S^{c}_{n}S^{c}_{\ell'}S^{a}_{n}S^{a}_{m}\Bigr\rangle}&=&\frac{1}{4}\delta^{ac}\Bigl\langle{S^{a}_{m}S^{c}_{\ell'}\Bigr\rangle}+\frac{i}{2}\epsilon_{cab}\Bigl\langle{S^{b}_{n}S^{a}_{m}S^{c}_{\ell'}\Bigr\rangle}. \nonumber \\ \label{eqn:2orderM_2}
\end{eqnarray}
The first term in each region gives a constant of $\tau', \tau''$, which does not  contribute to the integral in Eq.~(\ref{eqn:2orderM_1}). Now consider the second term with the Levi-Civita symbol, in a specific region, say $(1)$. Combining the 4 possibilities  marked by $\times$ in Fig.~\ref{fig:4spin_config} in region $(1)$, namely $n \ne m$ and (i): \{$\ell=n$; $\ell'\ne m,n$\}, (ii): \{$\ell=m$; $\ell'\ne m,n$\}, (iii): \{$\ell'=n$; $\ell\ne m,n$\}, (iv): \{$\ell'=m$; $\ell\ne m,n$\}, gives
\begin{eqnarray}
\Bigl\langle{S^{a}_{n}S^{a}_{m}S^{c}_{n}S^{c}_{\ell'}\Bigr\rangle}&=&\frac{1}{4}\delta^{ac}\Bigl\langle{S^{a}_{m}S^{c}_{\ell'}\Bigr\rangle}-\frac{i}{2}\epsilon_{cab}\Bigl\langle{S^{b}_{n}S^{a}_{m}S^{c}_{\ell'}\Bigr\rangle}, \nonumber \\ 
\Bigl\langle{S^{a}_{n}S^{a}_{m}S^{c}_{m}S^{c}_{\ell'}\Bigr\rangle}&=&\frac{1}{4}\delta^{ac}\Bigl\langle{S^{a}_{n}S^{c}_{\ell'}\Bigr\rangle}-\frac{i}{2}\epsilon_{cab}\Bigl\langle{S^{a}_{n}S^{b}_{m}S^{c}_{\ell'}\Bigr\rangle}, \nonumber \\ 
\Bigl\langle{S^{a}_{n}S^{a}_{m}S^{c}_{\ell'}S^{c}_{n}\Bigr\rangle}&=&\frac{1}{4}\delta^{ac}\Bigl\langle{S^{a}_{m}S^{c}_{\ell'}\Bigr\rangle}-\frac{i}{2}\epsilon_{cab}\Bigl\langle{S^{b}_{n}S^{a}_{m}S^{c}_{\ell'}\Bigr\rangle}, \nonumber \\ 
\Bigl\langle{S^{a}_{n}S^{a}_{m}S^{c}_{\ell'}S^{c}_{m}\Bigr\rangle}&=&\frac{1}{4}\delta^{ac}\Bigl\langle{S^{a}_{n}S^{c}_{\ell'}\Bigr\rangle}-\frac{i}{2}\epsilon_{cab}\Bigl\langle{S^{a}_{n}S^{b}_{m}S^{c}_{\ell'}\Bigr\rangle}, \nonumber \\ \label{eqn:2orderM_2aa}
\end{eqnarray}
Putting these configurations together, we see that all equally contribute, giving a relative factor of $-4$, depicted schematically in Fig.~\ref{fig:1pair}(b). This procedure can be repeated for regions $(2)$-$(6)$ to arrive at the remaining factors indicated in Fig.~\ref{fig:1pair}(b). We can then conclude, by symmetry of the integral under consideration, that the 4-spin correlator for a single pair of identical impurities must be null. 

This leaves the set of 4-spin correlators containing three and a double pair of identical impurities, as depicted in the boxed region of Fig.~\ref{fig:4spin_config}. In following the previously outlined procedure, the double pair of identical impurities results in a finite contribution, as depicted in Fig.~\ref{fig:2a3pair}(a). One needs to be careful in considering all possible configurations and their respective regions within the 4-spin correlator. For example, configuration \{$\ell=n$; $\ell'=m$\} in region $(1)$ simplifies as
\begin{eqnarray}
\sum_{c}\Bigl\langle{S^{a}_{n}S^{a}_{m}S^{c}_{n}S^{c}_{m}\Bigr\rangle}=\frac{1}{16}-\frac{1}{4}\sum_{c\ne a}\Bigl\langle{S^{c}_{n}S^{c}_{m}\Bigr\rangle}. \label{eqn:2orderM_2a}
\end{eqnarray}
All other regions generate the same result up to a sign of the second term. As in Eq.~(\ref{eqn:2orderM_2}), the first constant term cancels out. Utilizing Eq.~(\ref{eqn:asym_largek_5}), and up to the factors specified in Fig.~\ref{fig:2a3pair}(a) associated to the total polarity of the two contributing configurations, the relevant second-order contribution within each of these regions reduces from Eq.~(\ref{eqn:2orderM_1}) to 
\begin{eqnarray}
\braket{S^{a}_{n}(\tau_{n})S^{b}_{m}(\tau_{m})}_{(2)}&=&-\frac{k}{8}\lambda^{2}\Bigl(\frac{1}{6}j_{nm}(j_{nm}+1)-\frac{1}{4}\Bigr) \label{eqn:2orderM_3} \\ 
&&\times\int_{R_{1}} d\tau'd\tau''\frac{\delta^{ab}}{(\tau'-\tau''-i\Delta{x})^{2}}, \nonumber 
\end{eqnarray}
where $R_{1}$ refers to region $(1)$ in Fig.~\ref{fig:1pair}(a). 

Similarly, for the cases $\ell=\ell'=n\ne m$ and $\ell=\ell'=m\ne n$ depicted in Fig.~\ref{fig:2a3pair}(b), the 4-spin correlator is also finite. For $\ell=\ell'=n\ne m$ in region $(1)$, the 4-spin correlator reduces to
\begin{align}
\sum_{c}\Bigl\langle{S^{a}_{n}S^{a}_{m}S^{c}_{n}S^{c}_{n}\Bigr\rangle}=\frac{1}{4}\Bigl\langle{S^{a}_{n}S^{a}_{m}\Bigr\rangle}+\frac{1}{4}\sum_{c\ne a}\Bigl\langle{S^{c}_{n}S^{c}_{m}\Bigr\rangle}. \label{eqn:2orderM_3aa}
\end{align}
As before, only the second term contributes. Unlike the nonzero $\Delta{x}$ present in the previous configuration (i.e. see Eq.~(\ref{eqn:2orderM_3}) ), this integral contains an ultraviolet cutoff $\epsilon$,
\begin{eqnarray}
\braket{S^{a}_{n}(\tau_{n})S^{b}_{m}(\tau_{m})}_{(2)}&=&\frac{k}{8}\lambda^{2}\Bigl(\frac{1}{6}j_{nm}(j_{nm}+1)-\frac{1}{4}\Bigr) \label{eqn:2orderM_3a} \\ 
&&\times\int_{R_{1}}d\tau'd\tau''\frac{\delta^{ab}}{(\tau'-\tau''-i\epsilon)^{2}}. \nonumber
\end{eqnarray}
We can see the cancellation between Eq.~(\ref{eqn:2orderM_3}) and (\ref{eqn:2orderM_3a}) in Fig.~\ref{fig:2a3pair}. Similarly, we find the same cancellation in all regions but region $(4)$. 

\begin{widetext}

After adding the regions specified in figure~\ref{fig:2a3pair}(a) and \ref{fig:2a3pair}(b) together, the nonzero domain $\tau_{m}<\tau',\tau''<\tau_{n}$ depicted in figure~\ref{fig:2a3pair}(c) can be integrated to obtain a logarithmic trend which is independent of $\Delta{x}$ so long as $\tau_{n},\tau_{m}\gg\Delta{x}$

\begin{figure}[!ht]
	\raisebox{-\height}{\includegraphics[width=0.8\textwidth]{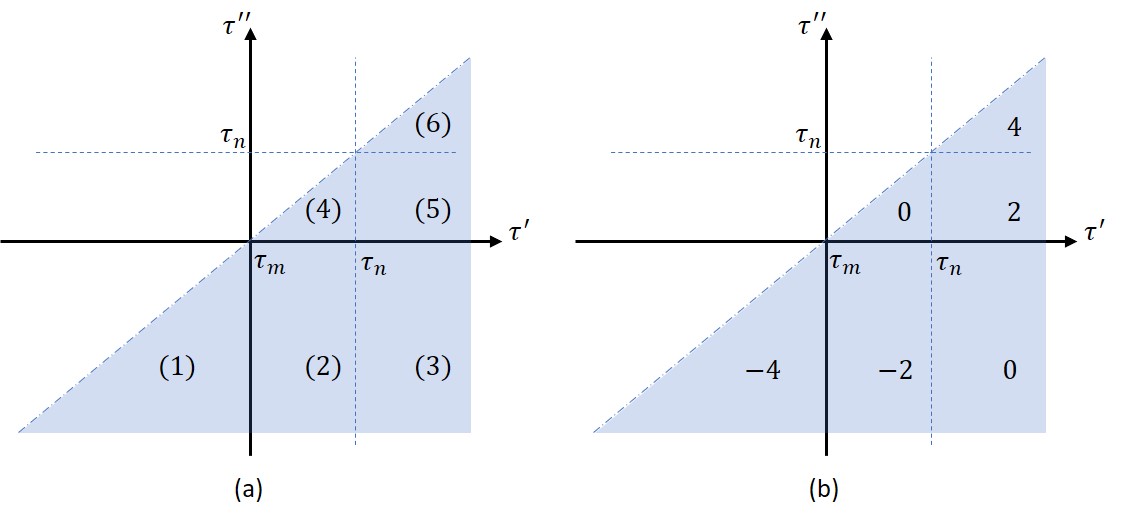}} 
	\caption{Time plot (a) indicates the different regions expressed in Eq.~(\ref{eqn:2orderM_2}). Time plot (b) specifies the factor accumulated in each regime for one pair of identical impurity-spins. By symmetry of the integral in Eq.~(\ref{eqn:2orderM_1}), the configurations marked by $\times$ in Fig.~\ref{fig:4spin_config} do not contribute.}
	\label{fig:1pair} 
\end{figure}

\begin{figure}[!ht]
	\raisebox{-\height}{\includegraphics[width=\textwidth]{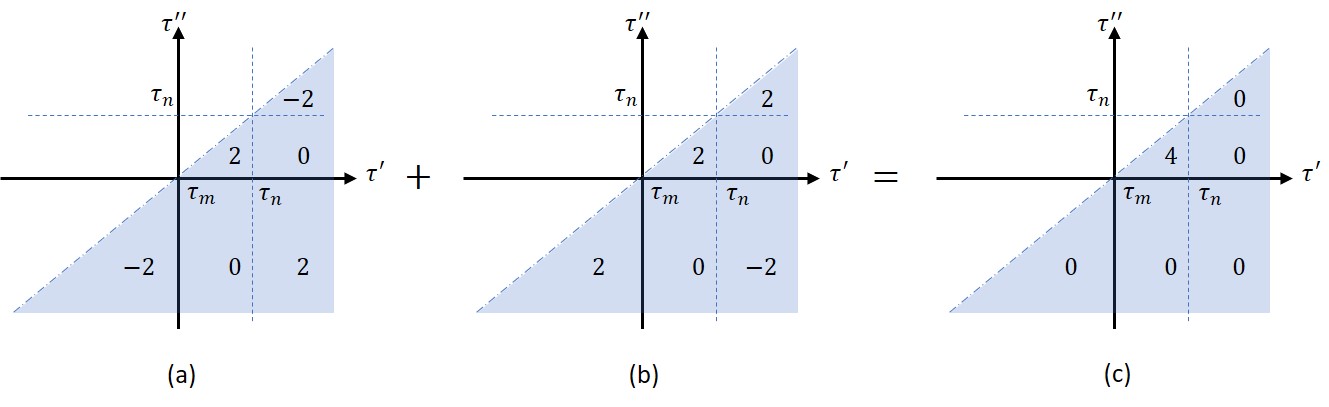}} 
	\caption{Time plots indicate the factors accumulated in each regime for configurations of (a) two pairs of identical impurity-spins and (b) three identical impurity-spins. These configurations are marked by a bold square in Fig.~\ref{fig:4spin_config}, with its cross diagonal and diagonal configurations corresponding to (a) and (b), respectively. Time plot (c) is the accumulative sum of factors within configurations (a) and (b).}
	\label{fig:2a3pair} 
\end{figure}

\begin{eqnarray}
I=\int_{\tau_{m}}^{\tau_{n}} d\tau' \int_{\tau_{m}}^{\tau'}d\tau'' \frac{1}{(\tau'-\tau''-i\Delta{x})^{2}}+\int_{\tau_{m}}^{\tau_{n}} d\tau'' \int_{\tau_{m}}^{\tau''}d\tau' \frac{1}{(\tau'-\tau''-i\Delta{x})^{2}}. \label{eqn:2orderM_6}
\end{eqnarray}
In switching the time coordinates $\tau'\leftrightarrow\tau''$ in the second integral, we have 
\begin{eqnarray}
I=I_{-}+I_{+}=\int_{\tau_{m}}^{\tau_{n}} d\tau' \int_{\tau_{m}}^{\tau'}d\tau'' \frac{1}{(\tau'-\tau''-i\Delta{x})^{2}}+\int_{\tau_{m}}^{\tau_{n}} d\tau' \int_{\tau_{m}}^{\tau'}d\tau'' \frac{1}{(\tau'-\tau''+i\Delta{x})^{2}}. \label{eqn:2orderM_7}
\end{eqnarray}
Evaluating each of these integrals gives
\begin{eqnarray}
I_{\pm}&=&\int_{\tau_{m}}^{\tau_{n}}d\tau' \int_{\tau_{m}}^{\tau'}d\tau'' \frac{1}{(\tau'-\tau''\pm i\Delta{x})^{2}} \\
&=&\int_{\tau_{m}}^{\tau_{n}}d\tau'\Bigl[\mp\frac{i}{\Delta{x}}-\frac{1}{\tau'-\tau_{m}\pm i\Delta{x}}\Bigr] \\
&=&-\log(\Delta{\tau}\pm i\Delta{x})+\log(\pm i\Delta{x})\mp\frac{i\Delta{\tau}}{\Delta{x}}. \label{eqn:2orderM_8}
\end{eqnarray}
Here, $\Delta{\tau}=\tau_{n}-\tau_{m}$. Plugging this back into Eq.~(\ref{eqn:2orderM_7}), with $\Delta\tau\gg\Delta{x}$, we obtain
\begin{eqnarray}
I=-2\Bigl[\log(\tau_{n}-\tau_{m})-\log(\Delta{x})\Bigr], \label{eqn:2orderM_9}
\end{eqnarray}
where the logarithmic divergence for small $\Delta{x}\to0$ can be renormalized, as in the one-impurity case. Furthermore, whether we consider $\Delta{x}$ or $\epsilon$, the above results do not change. We are left with a logarithmic dependence in time for the second-order perturbation of a multi-impurity correlator. 

Taking the result of Eq.~(\ref{eqn:2orderM_9}), accounting for the accumulated factor of $\times4$ from the different configurations (i.e. see Fig.~\ref{fig:2a3pair}(c) ), and letting $\lambda$ trend as $2/k$, the zeroth- and second-order contributions in Eq.~(\ref{eqn:Ispin_corrM}) are correctly obtained.

\end{widetext}

%

\begin{thebibliography}{52}%
	\makeatletter
	\providecommand \@ifxundefined [1]{%
		\@ifx{#1\undefined}
	}%
	\providecommand \@ifnum [1]{%
		\ifnum #1\expandafter \@firstoftwo
		\else \expandafter \@secondoftwo
		\fi
	}%
	\providecommand \@ifx [1]{%
		\ifx #1\expandafter \@firstoftwo
		\else \expandafter \@secondoftwo
		\fi
	}%
	\providecommand \natexlab [1]{#1}%
	\providecommand \enquote  [1]{``#1''}%
	\providecommand \bibnamefont  [1]{#1}%
	\providecommand \bibfnamefont [1]{#1}%
	\providecommand \citenamefont [1]{#1}%
	\providecommand \href@noop [0]{\@secondoftwo}%
	\providecommand \href [0]{\begingroup \@sanitize@url \@href}%
	\providecommand \@href[1]{\@@startlink{#1}\@@href}%
	\providecommand \@@href[1]{\endgroup#1\@@endlink}%
	\providecommand \@sanitize@url [0]{\catcode `\\12\catcode `\$12\catcode
		`\&12\catcode `\#12\catcode `\^12\catcode `\_12\catcode `\%12\relax}%
	\providecommand \@@startlink[1]{}%
	\providecommand \@@endlink[0]{}%
	\providecommand \url  [0]{\begingroup\@sanitize@url \@url }%
	\providecommand \@url [1]{\endgroup\@href {#1}{\urlprefix }}%
	\providecommand \urlprefix  [0]{URL }%
	\providecommand \Eprint [0]{\href }%
	\providecommand \doibase [0]{http://dx.doi.org/}%
	\providecommand \selectlanguage [0]{\@gobble}%
	\providecommand \bibinfo  [0]{\@secondoftwo}%
	\providecommand \bibfield  [0]{\@secondoftwo}%
	\providecommand \translation [1]{[#1]}%
	\providecommand \BibitemOpen [0]{}%
	\providecommand \bibitemStop [0]{}%
	\providecommand \bibitemNoStop [0]{.\EOS\space}%
	\providecommand \EOS [0]{\spacefactor3000\relax}%
	\providecommand \BibitemShut  [1]{\csname bibitem#1\endcsname}%
	\let\auto@bib@innerbib\@empty
	\bibitem [{\citenamefont {Affleck}\ and\ \citenamefont
		{Ludwig}(1993)}]{affleck1993exact}%
	\BibitemOpen
	\bibfield  {author} {\bibinfo {author} {\bibfnamefont {I.}~\bibnamefont
			{Affleck}}\ and\ \bibinfo {author} {\bibfnamefont {A.~W.}\ \bibnamefont
			{Ludwig}},\ }\href@noop {} {\bibfield  {journal} {\bibinfo  {journal} {Phys.
				Rev. B}\ }\textbf {\bibinfo {volume} {48}},\ \bibinfo {pages} {7297}
		(\bibinfo {year} {1993})}\BibitemShut {NoStop}%
	\bibitem [{\citenamefont {Ludwig}\ and\ \citenamefont
		{Affleck}(1994)}]{ludwig1994exact}%
	\BibitemOpen
	\bibfield  {author} {\bibinfo {author} {\bibfnamefont {A.~W.}\ \bibnamefont
			{Ludwig}}\ and\ \bibinfo {author} {\bibfnamefont {I.}~\bibnamefont
			{Affleck}},\ }\href@noop {} {\bibfield  {journal} {\bibinfo  {journal} {Nucl.
				Phys. B}\ }\textbf {\bibinfo {volume} {428}},\ \bibinfo {pages} {545}
		(\bibinfo {year} {1994})}\BibitemShut {NoStop}%
	\bibitem [{\citenamefont {Potok}\ \emph {et~al.}(2007)\citenamefont {Potok},
		\citenamefont {Rau}, \citenamefont {Shtrikman}, \citenamefont {Oreg},\ and\
		\citenamefont {Goldhaber-Gordon}}]{potok2007observation}%
	\BibitemOpen
	\bibfield  {author} {\bibinfo {author} {\bibfnamefont {R.}~\bibnamefont
			{Potok}}, \bibinfo {author} {\bibfnamefont {I.}~\bibnamefont {Rau}}, \bibinfo
		{author} {\bibfnamefont {H.}~\bibnamefont {Shtrikman}}, \bibinfo {author}
		{\bibfnamefont {Y.}~\bibnamefont {Oreg}}, \ and\ \bibinfo {author}
		{\bibfnamefont {D.}~\bibnamefont {Goldhaber-Gordon}},\ }\href@noop {}
	{\bibfield  {journal} {\bibinfo  {journal} {Nature}\ }\textbf {\bibinfo
			{volume} {446}},\ \bibinfo {pages} {167} (\bibinfo {year}
		{2007})}\BibitemShut {NoStop}%
	\bibitem [{\citenamefont {Andrei}(1980)}]{andrei1980diagonalization}%
	\BibitemOpen
	\bibfield  {author} {\bibinfo {author} {\bibfnamefont {N.}~\bibnamefont
			{Andrei}},\ }\href@noop {} {\bibfield  {journal} {\bibinfo  {journal} {Phys.
				Rev. Lett.}\ }\textbf {\bibinfo {volume} {45}},\ \bibinfo {pages} {379}
		(\bibinfo {year} {1980})}\BibitemShut {NoStop}%
	\bibitem [{\citenamefont {Vigman}(1980)}]{vigman1980exact}%
	\BibitemOpen
	\bibfield  {author} {\bibinfo {author} {\bibfnamefont {P.}~\bibnamefont
			{Vigman}},\ }\href@noop {} {\bibfield  {journal} {\bibinfo  {journal} {JETP
				Lett}\ }\textbf {\bibinfo {volume} {31}} (\bibinfo {year}
		{1980})}\BibitemShut {NoStop}%
	\bibitem [{\citenamefont {Affleck}\ and\ \citenamefont
		{Ludwig}(1991)}]{affleck1991universal}%
	\BibitemOpen
	\bibfield  {author} {\bibinfo {author} {\bibfnamefont {I.}~\bibnamefont
			{Affleck}}\ and\ \bibinfo {author} {\bibfnamefont {A.~W.}\ \bibnamefont
			{Ludwig}},\ }\href@noop {} {\bibfield  {journal} {\bibinfo  {journal} {Phys.
				Rev. Lett.}\ }\textbf {\bibinfo {volume} {67}},\ \bibinfo {pages} {161}
		(\bibinfo {year} {1991})}\BibitemShut {NoStop}%
	\bibitem [{\citenamefont {Han}\ \emph {et~al.}()\citenamefont {Han},
		\citenamefont {Mitchell}, \citenamefont {Iftikhar}, \citenamefont {Kleeorin},
		\citenamefont {Anthore}, \citenamefont {Pierre}, \citenamefont {Meir},\ and\
		\citenamefont {Sela}}]{han2021entropy}%
	\BibitemOpen
	\bibfield  {author} {\bibinfo {author} {\bibfnamefont {C.}~\bibnamefont
			{Han}}, \bibinfo {author} {\bibfnamefont {A.~K.}\ \bibnamefont {Mitchell}},
		\bibinfo {author} {\bibfnamefont {Z.}~\bibnamefont {Iftikhar}}, \bibinfo
		{author} {\bibfnamefont {Y.}~\bibnamefont {Kleeorin}}, \bibinfo {author}
		{\bibfnamefont {A.}~\bibnamefont {Anthore}}, \bibinfo {author} {\bibfnamefont
			{F.}~\bibnamefont {Pierre}}, \bibinfo {author} {\bibfnamefont
			{Y.}~\bibnamefont {Meir}}, \ and\ \bibinfo {author} {\bibfnamefont
			{E.}~\bibnamefont {Sela}},\ }\href@noop {} {\ }\Eprint
	{http://arxiv.org/abs/arXiv:2108.12878v1} {arXiv:2108.12878v1} \BibitemShut
	{NoStop}%
	\bibitem [{\citenamefont {Lopes}\ \emph {et~al.}(2020)\citenamefont {Lopes},
		\citenamefont {Affleck},\ and\ \citenamefont {Sela}}]{lopes2020anyons}%
	\BibitemOpen
	\bibfield  {author} {\bibinfo {author} {\bibfnamefont {P.~L.}\ \bibnamefont
			{Lopes}}, \bibinfo {author} {\bibfnamefont {I.}~\bibnamefont {Affleck}}, \
		and\ \bibinfo {author} {\bibfnamefont {E.}~\bibnamefont {Sela}},\ }\href@noop
	{} {\bibfield  {journal} {\bibinfo  {journal} {Phys. Rev. B}\ }\textbf
		{\bibinfo {volume} {101}},\ \bibinfo {pages} {085141} (\bibinfo {year}
		{2020})}\BibitemShut {NoStop}%
	\bibitem [{\citenamefont {Komijani}(2020)}]{komijani2020isolating}%
	\BibitemOpen
	\bibfield  {author} {\bibinfo {author} {\bibfnamefont {Y.}~\bibnamefont
			{Komijani}},\ }\href@noop {} {\bibfield  {journal} {\bibinfo  {journal}
			{Phys. Rev. B}\ }\textbf {\bibinfo {volume} {101}},\ \bibinfo {pages}
		{235131} (\bibinfo {year} {2020})}\BibitemShut {NoStop}%
	\bibitem [{\citenamefont {Nayak}\ \emph {et~al.}(2008)\citenamefont {Nayak},
		\citenamefont {Simon}, \citenamefont {Stern}, \citenamefont {Freedman},\ and\
		\citenamefont {Sarma}}]{nayak2008non}%
	\BibitemOpen
	\bibfield  {author} {\bibinfo {author} {\bibfnamefont {C.}~\bibnamefont
			{Nayak}}, \bibinfo {author} {\bibfnamefont {S.~H.}\ \bibnamefont {Simon}},
		\bibinfo {author} {\bibfnamefont {A.}~\bibnamefont {Stern}}, \bibinfo
		{author} {\bibfnamefont {M.}~\bibnamefont {Freedman}}, \ and\ \bibinfo
		{author} {\bibfnamefont {S.~D.}\ \bibnamefont {Sarma}},\ }\href@noop {}
	{\bibfield  {journal} {\bibinfo  {journal} {Rev. Mod. Phys.}\ }\textbf
		{\bibinfo {volume} {80}},\ \bibinfo {pages} {1083} (\bibinfo {year}
		{2008})}\BibitemShut {NoStop}%
	\bibitem [{\citenamefont {Bonderson}\ \emph {et~al.}(2008)\citenamefont
		{Bonderson}, \citenamefont {Freedman},\ and\ \citenamefont
		{Nayak}}]{bonderson2008measurement}%
	\BibitemOpen
	\bibfield  {author} {\bibinfo {author} {\bibfnamefont {P.}~\bibnamefont
			{Bonderson}}, \bibinfo {author} {\bibfnamefont {M.}~\bibnamefont {Freedman}},
		\ and\ \bibinfo {author} {\bibfnamefont {C.}~\bibnamefont {Nayak}},\
	}\href@noop {} {\bibfield  {journal} {\bibinfo  {journal} {Phys. Rev. Lett.}\
		}\textbf {\bibinfo {volume} {101}},\ \bibinfo {pages} {010501} (\bibinfo
		{year} {2008})}\BibitemShut {NoStop}%
	\bibitem [{\citenamefont {Kitaev}(2006)}]{kitaev2006anyons}%
	\BibitemOpen
	\bibfield  {author} {\bibinfo {author} {\bibfnamefont {A.}~\bibnamefont
			{Kitaev}},\ }\href@noop {} {\bibfield  {journal} {\bibinfo  {journal} {Ann.
				Phys.}\ }\textbf {\bibinfo {volume} {321}},\ \bibinfo {pages} {2} (\bibinfo
		{year} {2006})}\BibitemShut {NoStop}%
	\bibitem [{\citenamefont {Bonderson}(2012)}]{bonderson2012non}%
	\BibitemOpen
	\bibfield  {author} {\bibinfo {author} {\bibfnamefont {P.~H.}\ \bibnamefont
			{Bonderson}},\ }\href@noop {} {\emph {\bibinfo {title} {Non-Abelian anyons
				and interferometry}}}\ (\bibinfo  {publisher} {California Institute of
		Technology},\ \bibinfo {year} {2012})\BibitemShut {NoStop}%
	\bibitem [{Note1()}]{Note1}%
	\BibitemOpen
	\bibinfo {note} {We take the chance to remark here that Chern insulating
		systems would, in fact, display less difficulties, with geometrically
		overlapping channels and no need for magnetic fields to define chiral
		channels}\BibitemShut {NoStop}%
	\bibitem [{\citenamefont {Emery}\ and\ \citenamefont
		{Kivelson}(1992)}]{emery1992mapping}%
	\BibitemOpen
	\bibfield  {author} {\bibinfo {author} {\bibfnamefont {V.}~\bibnamefont
			{Emery}}\ and\ \bibinfo {author} {\bibfnamefont {S.}~\bibnamefont
			{Kivelson}},\ }\href@noop {} {\bibfield  {journal} {\bibinfo  {journal}
			{Phys. Rev. B}\ }\textbf {\bibinfo {volume} {46}},\ \bibinfo {pages} {10812}
		(\bibinfo {year} {1992})}\BibitemShut {NoStop}%
	\bibitem [{inP()}]{inPrepLotem}%
	\BibitemOpen
	\href@noop {} {}\bibinfo {note} {M.~Lotem \emph{et. al., ~in
			preparation}.}\BibitemShut {Stop}%
	\bibitem [{\citenamefont {Cardy}(1984)}]{CARDY_1}%
	\BibitemOpen
	\bibfield  {author} {\bibinfo {author} {\bibfnamefont {J.~L.}\ \bibnamefont
			{Cardy}},\ }\href {\doibase https://doi.org/10.1016/0550-3213(84)90241-4}
	{\bibfield  {journal} {\bibinfo  {journal} {Nucl. Phys. B}\ }\textbf
		{\bibinfo {volume} {240}},\ \bibinfo {pages} {514 } (\bibinfo {year}
		{1984})}\BibitemShut {NoStop}%
	\bibitem [{\citenamefont {Cardy}(1989)}]{CARDY_2}%
	\BibitemOpen
	\bibfield  {author} {\bibinfo {author} {\bibfnamefont {J.~L.}\ \bibnamefont
			{Cardy}},\ }\href {\doibase https://doi.org/10.1016/0550-3213(89)90521-X}
	{\bibfield  {journal} {\bibinfo  {journal} {Nucl. Phys. B}\ }\textbf
		{\bibinfo {volume} {324}},\ \bibinfo {pages} {581 } (\bibinfo {year}
		{1989})}\BibitemShut {NoStop}%
	\bibitem [{\citenamefont {Cardy}\ and\ \citenamefont
		{Lewellen}(1991)}]{Cardy_Lewellen}%
	\BibitemOpen
	\bibfield  {author} {\bibinfo {author} {\bibfnamefont {J.~L.}\ \bibnamefont
			{Cardy}}\ and\ \bibinfo {author} {\bibfnamefont {D.~C.}\ \bibnamefont
			{Lewellen}},\ }\href {\doibase https://doi.org/10.1016/0370-2693(91)90828-E}
	{\bibfield  {journal} {\bibinfo  {journal} {Phys. Lett. B}\ }\textbf
		{\bibinfo {volume} {259}},\ \bibinfo {pages} {274 } (\bibinfo {year}
		{1991})}\BibitemShut {NoStop}%
	\bibitem [{\citenamefont {Cardy}(2006)}]{Cardy_review}%
	\BibitemOpen
	\bibfield  {author} {\bibinfo {author} {\bibfnamefont {J.}~\bibnamefont
			{Cardy}},\ }in\ \href {\doibase
		https://doi.org/10.1016/B0-12-512666-2/00398-9} {\emph {\bibinfo {booktitle}
			{Encyclopedia of Mathematical Physics}}},\ \bibinfo {editor} {edited by\
		\bibinfo {editor} {\bibfnamefont {J.-P.}\ \bibnamefont {Francoise}}, \bibinfo
		{editor} {\bibfnamefont {G.~L.}\ \bibnamefont {Naber}}, \ and\ \bibinfo
		{editor} {\bibfnamefont {T.~S.}\ \bibnamefont {Tsun}}}\ (\bibinfo
	{publisher} {Academic Press},\ \bibinfo {address} {Oxford},\ \bibinfo {year}
	{2006})\ pp.\ \bibinfo {pages} {333 -- 340}\BibitemShut {NoStop}%
	\bibitem [{\citenamefont {Affleck}\ \emph {et~al.}(1992)\citenamefont
		{Affleck}, \citenamefont {Ludwig}, \citenamefont {Pang},\ and\ \citenamefont
		{Cox}}]{affleck1992relevance}%
	\BibitemOpen
	\bibfield  {author} {\bibinfo {author} {\bibfnamefont {I.}~\bibnamefont
			{Affleck}}, \bibinfo {author} {\bibfnamefont {A.~W.}\ \bibnamefont {Ludwig}},
		\bibinfo {author} {\bibfnamefont {H.-B.}\ \bibnamefont {Pang}}, \ and\
		\bibinfo {author} {\bibfnamefont {D.}~\bibnamefont {Cox}},\ }\href@noop {}
	{\bibfield  {journal} {\bibinfo  {journal} {Phys. Rev. B}\ }\textbf {\bibinfo
			{volume} {45}},\ \bibinfo {pages} {7918} (\bibinfo {year}
		{1992})}\BibitemShut {NoStop}%
	\bibitem [{\citenamefont {Pustilnik}\ \emph {et~al.}(2004)\citenamefont
		{Pustilnik}, \citenamefont {Borda}, \citenamefont {Glazman},\ and\
		\citenamefont {Von~Delft}}]{pustilnik2004quantum}%
	\BibitemOpen
	\bibfield  {author} {\bibinfo {author} {\bibfnamefont {M.}~\bibnamefont
			{Pustilnik}}, \bibinfo {author} {\bibfnamefont {L.}~\bibnamefont {Borda}},
		\bibinfo {author} {\bibfnamefont {L.}~\bibnamefont {Glazman}}, \ and\
		\bibinfo {author} {\bibfnamefont {J.}~\bibnamefont {Von~Delft}},\ }\href@noop
	{} {\bibfield  {journal} {\bibinfo  {journal} {Phys. Rev. B}\ }\textbf
		{\bibinfo {volume} {69}},\ \bibinfo {pages} {115316} (\bibinfo {year}
		{2004})}\BibitemShut {NoStop}%
	\bibitem [{\citenamefont {Sela}\ \emph {et~al.}(2011)\citenamefont {Sela},
		\citenamefont {Mitchell},\ and\ \citenamefont {Fritz}}]{sela2011exact}%
	\BibitemOpen
	\bibfield  {author} {\bibinfo {author} {\bibfnamefont {E.}~\bibnamefont
			{Sela}}, \bibinfo {author} {\bibfnamefont {A.~K.}\ \bibnamefont {Mitchell}},
		\ and\ \bibinfo {author} {\bibfnamefont {L.}~\bibnamefont {Fritz}},\ }\href
	{\doibase 10.1103/PhysRevLett.106.147202} {\bibfield  {journal} {\bibinfo
			{journal} {Phys. Rev. Lett.}\ }\textbf {\bibinfo {volume} {106}},\ \bibinfo
		{pages} {147202} (\bibinfo {year} {2011})}\BibitemShut {NoStop}%
	\bibitem [{\citenamefont {Mitchell}\ and\ \citenamefont
		{Sela}(2012)}]{mitchell2012universal}%
	\BibitemOpen
	\bibfield  {author} {\bibinfo {author} {\bibfnamefont {A.~K.}\ \bibnamefont
			{Mitchell}}\ and\ \bibinfo {author} {\bibfnamefont {E.}~\bibnamefont
			{Sela}},\ }\href@noop {} {\bibfield  {journal} {\bibinfo  {journal} {Phys.
				Rev. B}\ }\textbf {\bibinfo {volume} {85}},\ \bibinfo {pages} {235127}
		(\bibinfo {year} {2012})}\BibitemShut {NoStop}%
	\bibitem [{\citenamefont {Keller}\ \emph {et~al.}(2015)\citenamefont {Keller},
		\citenamefont {Peeters}, \citenamefont {Moca}, \citenamefont {Weymann},
		\citenamefont {Mahalu}, \citenamefont {Umansky}, \citenamefont {Zar{\'a}nd},\
		and\ \citenamefont {Goldhaber-Gordon}}]{keller2015universal}%
	\BibitemOpen
	\bibfield  {author} {\bibinfo {author} {\bibfnamefont {A.}~\bibnamefont
			{Keller}}, \bibinfo {author} {\bibfnamefont {L.}~\bibnamefont {Peeters}},
		\bibinfo {author} {\bibfnamefont {C.}~\bibnamefont {Moca}}, \bibinfo {author}
		{\bibfnamefont {I.}~\bibnamefont {Weymann}}, \bibinfo {author} {\bibfnamefont
			{D.}~\bibnamefont {Mahalu}}, \bibinfo {author} {\bibfnamefont
			{V.}~\bibnamefont {Umansky}}, \bibinfo {author} {\bibfnamefont
			{G.}~\bibnamefont {Zar{\'a}nd}}, \ and\ \bibinfo {author} {\bibfnamefont
			{D.}~\bibnamefont {Goldhaber-Gordon}},\ }\href@noop {} {\bibfield  {journal}
		{\bibinfo  {journal} {Nature}\ }\textbf {\bibinfo {volume} {526}},\ \bibinfo
		{pages} {237} (\bibinfo {year} {2015})}\BibitemShut {NoStop}%
	\bibitem [{\citenamefont {Nozieres}\ and\ \citenamefont
		{Blandin}(1980)}]{nozieres1980kondo}%
	\BibitemOpen
	\bibfield  {author} {\bibinfo {author} {\bibfnamefont {P.}~\bibnamefont
			{Nozieres}}\ and\ \bibinfo {author} {\bibfnamefont {A.}~\bibnamefont
			{Blandin}},\ }\href@noop {} {\bibfield  {journal} {\bibinfo  {journal} {J.
				Phys.}\ }\textbf {\bibinfo {volume} {41}},\ \bibinfo {pages} {193} (\bibinfo
		{year} {1980})}\BibitemShut {NoStop}%
	\bibitem [{\citenamefont {Von~Delft}\ and\ \citenamefont
		{Schoeller}(1998)}]{delft1998bosonize}%
	\BibitemOpen
	\bibfield  {author} {\bibinfo {author} {\bibfnamefont {J.}~\bibnamefont
			{Von~Delft}}\ and\ \bibinfo {author} {\bibfnamefont {H.}~\bibnamefont
			{Schoeller}},\ }\href@noop {} {\bibfield  {journal} {\bibinfo  {journal}
			{Ann. Phys.}\ }\textbf {\bibinfo {volume} {7}},\ \bibinfo {pages} {225}
		(\bibinfo {year} {1998})}\BibitemShut {NoStop}%
	\bibitem [{\citenamefont {von Delft}\ \emph {et~al.}(1998)\citenamefont {von
			Delft}, \citenamefont {Zar\'and},\ and\ \citenamefont
		{Fabrizio}}]{delft1998kondo}%
	\BibitemOpen
	\bibfield  {author} {\bibinfo {author} {\bibfnamefont {J.}~\bibnamefont {von
				Delft}}, \bibinfo {author} {\bibfnamefont {G.}~\bibnamefont {Zar\'and}}, \
		and\ \bibinfo {author} {\bibfnamefont {M.}~\bibnamefont {Fabrizio}},\ }\href
	{\doibase 10.1103/PhysRevLett.81.196} {\bibfield  {journal} {\bibinfo
			{journal} {Phys. Rev. Lett.}\ }\textbf {\bibinfo {volume} {81}},\ \bibinfo
		{pages} {196} (\bibinfo {year} {1998})}\BibitemShut {NoStop}%
	\bibitem [{\citenamefont {Sela}\ and\ \citenamefont
		{Affleck}(2009{\natexlab{a}})}]{sela2009nonequilibrium1}%
	\BibitemOpen
	\bibfield  {author} {\bibinfo {author} {\bibfnamefont {E.}~\bibnamefont
			{Sela}}\ and\ \bibinfo {author} {\bibfnamefont {I.}~\bibnamefont {Affleck}},\
	}\href@noop {} {\bibfield  {journal} {\bibinfo  {journal} {Phys. Rev. Lett.}\
		}\textbf {\bibinfo {volume} {102}},\ \bibinfo {pages} {047201} (\bibinfo
		{year} {2009}{\natexlab{a}})}\BibitemShut {NoStop}%
	\bibitem [{\citenamefont {Sela}\ and\ \citenamefont
		{Affleck}(2009{\natexlab{b}})}]{sela2009nonequilibrium2}%
	\BibitemOpen
	\bibfield  {author} {\bibinfo {author} {\bibfnamefont {E.}~\bibnamefont
			{Sela}}\ and\ \bibinfo {author} {\bibfnamefont {I.}~\bibnamefont {Affleck}},\
	}\href@noop {} {\bibfield  {journal} {\bibinfo  {journal} {Phys. Rev. B}\
		}\textbf {\bibinfo {volume} {79}},\ \bibinfo {pages} {125110} (\bibinfo
		{year} {2009}{\natexlab{b}})}\BibitemShut {NoStop}%
	\bibitem [{\citenamefont {Landau}\ \emph {et~al.}(2018)\citenamefont {Landau},
		\citenamefont {Cornfeld},\ and\ \citenamefont {Sela}}]{landau2018charge}%
	\BibitemOpen
	\bibfield  {author} {\bibinfo {author} {\bibfnamefont {L.~A.}\ \bibnamefont
			{Landau}}, \bibinfo {author} {\bibfnamefont {E.}~\bibnamefont {Cornfeld}}, \
		and\ \bibinfo {author} {\bibfnamefont {E.}~\bibnamefont {Sela}},\ }\href@noop
	{} {\bibfield  {journal} {\bibinfo  {journal} {Phys. Rev. Lett.}\ }\textbf
		{\bibinfo {volume} {120}},\ \bibinfo {pages} {186801} (\bibinfo {year}
		{2018})}\BibitemShut {NoStop}%
	\bibitem [{\citenamefont {Kitaev}(2001)}]{kitaev2001unpaired}%
	\BibitemOpen
	\bibfield  {author} {\bibinfo {author} {\bibfnamefont {A.~Y.}\ \bibnamefont
			{Kitaev}},\ }\href@noop {} {\bibfield  {journal} {\bibinfo  {journal}
			{Phys.-Uspekhi}\ }\textbf {\bibinfo {volume} {44}},\ \bibinfo {pages} {131}
		(\bibinfo {year} {2001})}\BibitemShut {NoStop}%
	\bibitem [{\citenamefont {Alicea}(2012)}]{alicea2012new}%
	\BibitemOpen
	\bibfield  {author} {\bibinfo {author} {\bibfnamefont {J.}~\bibnamefont
			{Alicea}},\ }\href@noop {} {\bibfield  {journal} {\bibinfo  {journal} {Rep.
				Prog. Phys.}\ }\textbf {\bibinfo {volume} {75}},\ \bibinfo {pages} {076501}
		(\bibinfo {year} {2012})}\BibitemShut {NoStop}%
	\bibitem [{\citenamefont {Leijnse}\ and\ \citenamefont
		{Flensberg}(2012)}]{leijnse2012introduction}%
	\BibitemOpen
	\bibfield  {author} {\bibinfo {author} {\bibfnamefont {M.}~\bibnamefont
			{Leijnse}}\ and\ \bibinfo {author} {\bibfnamefont {K.}~\bibnamefont
			{Flensberg}},\ }\href@noop {} {\bibfield  {journal} {\bibinfo  {journal}
			{Semicond. Sci. Technol.}\ }\textbf {\bibinfo {volume} {27}},\ \bibinfo
		{pages} {124003} (\bibinfo {year} {2012})}\BibitemShut {NoStop}%
	\bibitem [{\citenamefont {Beenakker}(2013)}]{beenakker2013search}%
	\BibitemOpen
	\bibfield  {author} {\bibinfo {author} {\bibfnamefont {C.}~\bibnamefont
			{Beenakker}},\ }\href@noop {} {\bibfield  {journal} {\bibinfo  {journal}
			{Annu. Rev. Condens. Matter Phys.}\ }\textbf {\bibinfo {volume} {4}},\
		\bibinfo {pages} {113} (\bibinfo {year} {2013})}\BibitemShut {NoStop}%
	\bibitem [{\citenamefont {Mourik}\ \emph {et~al.}(2012)\citenamefont {Mourik},
		\citenamefont {Zuo}, \citenamefont {Frolov}, \citenamefont {Plissard},
		\citenamefont {Bakkers},\ and\ \citenamefont
		{Kouwenhoven}}]{mourik2012signatures}%
	\BibitemOpen
	\bibfield  {author} {\bibinfo {author} {\bibfnamefont {V.}~\bibnamefont
			{Mourik}}, \bibinfo {author} {\bibfnamefont {K.}~\bibnamefont {Zuo}},
		\bibinfo {author} {\bibfnamefont {S.~M.}\ \bibnamefont {Frolov}}, \bibinfo
		{author} {\bibfnamefont {S.}~\bibnamefont {Plissard}}, \bibinfo {author}
		{\bibfnamefont {E.~P.}\ \bibnamefont {Bakkers}}, \ and\ \bibinfo {author}
		{\bibfnamefont {L.~P.}\ \bibnamefont {Kouwenhoven}},\ }\href@noop {}
	{\bibfield  {journal} {\bibinfo  {journal} {Science}\ }\textbf {\bibinfo
			{volume} {336}},\ \bibinfo {pages} {1003} (\bibinfo {year}
		{2012})}\BibitemShut {NoStop}%
	\bibitem [{\citenamefont {Das}\ \emph {et~al.}(2012)\citenamefont {Das},
		\citenamefont {Ronen}, \citenamefont {Most}, \citenamefont {Oreg},
		\citenamefont {Heiblum},\ and\ \citenamefont {Shtrikman}}]{das2012zero}%
	\BibitemOpen
	\bibfield  {author} {\bibinfo {author} {\bibfnamefont {A.}~\bibnamefont
			{Das}}, \bibinfo {author} {\bibfnamefont {Y.}~\bibnamefont {Ronen}}, \bibinfo
		{author} {\bibfnamefont {Y.}~\bibnamefont {Most}}, \bibinfo {author}
		{\bibfnamefont {Y.}~\bibnamefont {Oreg}}, \bibinfo {author} {\bibfnamefont
			{M.}~\bibnamefont {Heiblum}}, \ and\ \bibinfo {author} {\bibfnamefont
			{H.}~\bibnamefont {Shtrikman}},\ }\href@noop {} {\bibfield  {journal}
		{\bibinfo  {journal} {Nat. Phys.}\ }\textbf {\bibinfo {volume} {8}},\
		\bibinfo {pages} {887} (\bibinfo {year} {2012})}\BibitemShut {NoStop}%
	\bibitem [{\citenamefont {Plugge}\ \emph {et~al.}(2016)\citenamefont {Plugge},
		\citenamefont {Landau}, \citenamefont {Sela}, \citenamefont {Altland},
		\citenamefont {Flensberg},\ and\ \citenamefont {Egger}}]{plugge2016roadmap}%
	\BibitemOpen
	\bibfield  {author} {\bibinfo {author} {\bibfnamefont {S.}~\bibnamefont
			{Plugge}}, \bibinfo {author} {\bibfnamefont {L.}~\bibnamefont {Landau}},
		\bibinfo {author} {\bibfnamefont {E.}~\bibnamefont {Sela}}, \bibinfo {author}
		{\bibfnamefont {A.}~\bibnamefont {Altland}}, \bibinfo {author} {\bibfnamefont
			{K.}~\bibnamefont {Flensberg}}, \ and\ \bibinfo {author} {\bibfnamefont
			{R.}~\bibnamefont {Egger}},\ }\href@noop {} {\bibfield  {journal} {\bibinfo
			{journal} {Phys. Rev. B}\ }\textbf {\bibinfo {volume} {94}},\ \bibinfo
		{pages} {174514} (\bibinfo {year} {2016})}\BibitemShut {NoStop}%
	\bibitem [{\citenamefont {Vijay}\ and\ \citenamefont
		{Fu}(2016)}]{vijay2016teleportation}%
	\BibitemOpen
	\bibfield  {author} {\bibinfo {author} {\bibfnamefont {S.}~\bibnamefont
			{Vijay}}\ and\ \bibinfo {author} {\bibfnamefont {L.}~\bibnamefont {Fu}},\
	}\href@noop {} {\bibfield  {journal} {\bibinfo  {journal} {Phys. Rev. B}\
		}\textbf {\bibinfo {volume} {94}},\ \bibinfo {pages} {235446} (\bibinfo
		{year} {2016})}\BibitemShut {NoStop}%
	\bibitem [{\citenamefont {Dahan}\ \emph {et~al.}(2017)\citenamefont {Dahan},
		\citenamefont {Ahari}, \citenamefont {Ortiz}, \citenamefont {Seradjeh},\ and\
		\citenamefont {Grosfeld}}]{dahan2017non}%
	\BibitemOpen
	\bibfield  {author} {\bibinfo {author} {\bibfnamefont {D.}~\bibnamefont
			{Dahan}}, \bibinfo {author} {\bibfnamefont {M.~T.}\ \bibnamefont {Ahari}},
		\bibinfo {author} {\bibfnamefont {G.}~\bibnamefont {Ortiz}}, \bibinfo
		{author} {\bibfnamefont {B.}~\bibnamefont {Seradjeh}}, \ and\ \bibinfo
		{author} {\bibfnamefont {E.}~\bibnamefont {Grosfeld}},\ }\href@noop {}
	{\bibfield  {journal} {\bibinfo  {journal} {Phys. Rev. B}\ }\textbf {\bibinfo
			{volume} {95}},\ \bibinfo {pages} {201114} (\bibinfo {year}
		{2017})}\BibitemShut {NoStop}%
	\bibitem [{\citenamefont {Dahan}\ \emph {et~al.}(2020)\citenamefont {Dahan},
		\citenamefont {Grosfeld},\ and\ \citenamefont
		{Seradjeh}}]{dahan2020dynamically}%
	\BibitemOpen
	\bibfield  {author} {\bibinfo {author} {\bibfnamefont {D.}~\bibnamefont
			{Dahan}}, \bibinfo {author} {\bibfnamefont {E.}~\bibnamefont {Grosfeld}}, \
		and\ \bibinfo {author} {\bibfnamefont {B.}~\bibnamefont {Seradjeh}},\
	}\href@noop {} {\bibfield  {journal} {\bibinfo  {journal} {Phys. Rev. B}\
		}\textbf {\bibinfo {volume} {102}},\ \bibinfo {pages} {125142} (\bibinfo
		{year} {2020})}\BibitemShut {NoStop}%
	\bibitem [{\citenamefont {Smirnov}(2015)}]{smirnov2015majorana}%
	\BibitemOpen
	\bibfield  {author} {\bibinfo {author} {\bibfnamefont {S.}~\bibnamefont
			{Smirnov}},\ }\href@noop {} {\bibfield  {journal} {\bibinfo  {journal}
			{Physical Review B}\ }\textbf {\bibinfo {volume} {92}},\ \bibinfo {pages}
		{195312} (\bibinfo {year} {2015})}\BibitemShut {NoStop}%
	\bibitem [{\citenamefont {Sela}\ \emph {et~al.}(2019)\citenamefont {Sela},
		\citenamefont {Oreg}, \citenamefont {Plugge}, \citenamefont {Hartman},
		\citenamefont {L{\"u}scher},\ and\ \citenamefont {Folk}}]{sela2019detecting}%
	\BibitemOpen
	\bibfield  {author} {\bibinfo {author} {\bibfnamefont {E.}~\bibnamefont
			{Sela}}, \bibinfo {author} {\bibfnamefont {Y.}~\bibnamefont {Oreg}}, \bibinfo
		{author} {\bibfnamefont {S.}~\bibnamefont {Plugge}}, \bibinfo {author}
		{\bibfnamefont {N.}~\bibnamefont {Hartman}}, \bibinfo {author} {\bibfnamefont
			{S.}~\bibnamefont {L{\"u}scher}}, \ and\ \bibinfo {author} {\bibfnamefont
			{J.}~\bibnamefont {Folk}},\ }\href@noop {} {\bibfield  {journal} {\bibinfo
			{journal} {Phys. Rev. Lett.}\ }\textbf {\bibinfo {volume} {123}},\ \bibinfo
		{pages} {147702} (\bibinfo {year} {2019})}\BibitemShut {NoStop}%
	\bibitem [{\citenamefont {Rainis}\ and\ \citenamefont
		{Loss}(2012)}]{rainis2012majorana}%
	\BibitemOpen
	\bibfield  {author} {\bibinfo {author} {\bibfnamefont {D.}~\bibnamefont
			{Rainis}}\ and\ \bibinfo {author} {\bibfnamefont {D.}~\bibnamefont {Loss}},\
	}\href@noop {} {\bibfield  {journal} {\bibinfo  {journal} {Phys. Rev. B}\
		}\textbf {\bibinfo {volume} {85}},\ \bibinfo {pages} {174533} (\bibinfo
		{year} {2012})}\BibitemShut {NoStop}%
	\bibitem [{\citenamefont {Albrecht}\ \emph {et~al.}(2017)\citenamefont
		{Albrecht}, \citenamefont {Hansen}, \citenamefont {Higginbotham},
		\citenamefont {Kuemmeth}, \citenamefont {Jespersen}, \citenamefont
		{Nyg{\aa}rd}, \citenamefont {Krogstrup}, \citenamefont {Danon}, \citenamefont
		{Flensberg},\ and\ \citenamefont {Marcus}}]{albrecht2017transport}%
	\BibitemOpen
	\bibfield  {author} {\bibinfo {author} {\bibfnamefont {S.}~\bibnamefont
			{Albrecht}}, \bibinfo {author} {\bibfnamefont {E.}~\bibnamefont {Hansen}},
		\bibinfo {author} {\bibfnamefont {A.~P.}\ \bibnamefont {Higginbotham}},
		\bibinfo {author} {\bibfnamefont {F.}~\bibnamefont {Kuemmeth}}, \bibinfo
		{author} {\bibfnamefont {T.}~\bibnamefont {Jespersen}}, \bibinfo {author}
		{\bibfnamefont {J.}~\bibnamefont {Nyg{\aa}rd}}, \bibinfo {author}
		{\bibfnamefont {P.}~\bibnamefont {Krogstrup}}, \bibinfo {author}
		{\bibfnamefont {J.}~\bibnamefont {Danon}}, \bibinfo {author} {\bibfnamefont
			{K.}~\bibnamefont {Flensberg}}, \ and\ \bibinfo {author} {\bibfnamefont
			{C.}~\bibnamefont {Marcus}},\ }\href@noop {} {\bibfield  {journal} {\bibinfo
			{journal} {Phys. Rev. Lett.}\ }\textbf {\bibinfo {volume} {118}},\ \bibinfo
		{pages} {137701} (\bibinfo {year} {2017})}\BibitemShut {NoStop}%
	\bibitem [{\citenamefont {Karzig}\ \emph {et~al.}(2021)\citenamefont {Karzig},
		\citenamefont {Cole},\ and\ \citenamefont
		{Pikulin}}]{karzig2021quasiparticle}%
	\BibitemOpen
	\bibfield  {author} {\bibinfo {author} {\bibfnamefont {T.}~\bibnamefont
			{Karzig}}, \bibinfo {author} {\bibfnamefont {W.~S.}\ \bibnamefont {Cole}}, \
		and\ \bibinfo {author} {\bibfnamefont {D.~I.}\ \bibnamefont {Pikulin}},\
	}\href@noop {} {\bibfield  {journal} {\bibinfo  {journal} {Phys. Rev. Lett.}\
		}\textbf {\bibinfo {volume} {126}},\ \bibinfo {pages} {057702} (\bibinfo
		{year} {2021})}\BibitemShut {NoStop}%
	\bibitem [{\citenamefont {Bonderson}\ \emph {et~al.}(2009)\citenamefont
		{Bonderson}, \citenamefont {Freedman},\ and\ \citenamefont
		{Nayak}}]{bonderson2009measurement}%
	\BibitemOpen
	\bibfield  {author} {\bibinfo {author} {\bibfnamefont {P.}~\bibnamefont
			{Bonderson}}, \bibinfo {author} {\bibfnamefont {M.}~\bibnamefont {Freedman}},
		\ and\ \bibinfo {author} {\bibfnamefont {C.}~\bibnamefont {Nayak}},\
	}\href@noop {} {\bibfield  {journal} {\bibinfo  {journal} {Ann. Phys.}\
		}\textbf {\bibinfo {volume} {324}},\ \bibinfo {pages} {787} (\bibinfo {year}
		{2009})}\BibitemShut {NoStop}%
	\bibitem [{\citenamefont {Ingersent}\ \emph {et~al.}(2005)\citenamefont
		{Ingersent}, \citenamefont {Ludwig},\ and\ \citenamefont
		{Affleck}}]{ingersent2005kondo}%
	\BibitemOpen
	\bibfield  {author} {\bibinfo {author} {\bibfnamefont {K.}~\bibnamefont
			{Ingersent}}, \bibinfo {author} {\bibfnamefont {A.~W.}\ \bibnamefont
			{Ludwig}}, \ and\ \bibinfo {author} {\bibfnamefont {I.}~\bibnamefont
			{Affleck}},\ }\href@noop {} {\bibfield  {journal} {\bibinfo  {journal} {Phys.
				Rev. Lett.}\ }\textbf {\bibinfo {volume} {95}},\ \bibinfo {pages} {257204}
		(\bibinfo {year} {2005})}\BibitemShut {NoStop}%
	\bibitem [{\citenamefont {Lopez}\ \emph {et~al.}(2013)\citenamefont {Lopez},
		\citenamefont {Rejec}, \citenamefont {Martinek} \emph {et~al.}}]{lopez20133}%
	\BibitemOpen
	\bibfield  {author} {\bibinfo {author} {\bibfnamefont {R.}~\bibnamefont
			{Lopez}}, \bibinfo {author} {\bibfnamefont {T.}~\bibnamefont {Rejec}},
		\bibinfo {author} {\bibfnamefont {J.}~\bibnamefont {Martinek}},  \emph
		{et~al.},\ }\href@noop {} {\bibfield  {journal} {\bibinfo  {journal} {Phys.
				Rev. B}\ }\textbf {\bibinfo {volume} {87}},\ \bibinfo {pages} {035135}
		(\bibinfo {year} {2013})}\BibitemShut {NoStop}%
	\bibitem [{\citenamefont {Keller}\ \emph {et~al.}(2014)\citenamefont {Keller},
		\citenamefont {Amasha}, \citenamefont {Weymann}, \citenamefont {Moca},
		\citenamefont {Rau}, \citenamefont {Katine}, \citenamefont {Shtrikman},
		\citenamefont {Zar{\'a}nd},\ and\ \citenamefont
		{Goldhaber-Gordon}}]{keller2014emergent}%
	\BibitemOpen
	\bibfield  {author} {\bibinfo {author} {\bibfnamefont {A.}~\bibnamefont
			{Keller}}, \bibinfo {author} {\bibfnamefont {S.}~\bibnamefont {Amasha}},
		\bibinfo {author} {\bibfnamefont {I.}~\bibnamefont {Weymann}}, \bibinfo
		{author} {\bibfnamefont {C.}~\bibnamefont {Moca}}, \bibinfo {author}
		{\bibfnamefont {I.}~\bibnamefont {Rau}}, \bibinfo {author} {\bibfnamefont
			{J.}~\bibnamefont {Katine}}, \bibinfo {author} {\bibfnamefont
			{H.}~\bibnamefont {Shtrikman}}, \bibinfo {author} {\bibfnamefont
			{G.}~\bibnamefont {Zar{\'a}nd}}, \ and\ \bibinfo {author} {\bibfnamefont
			{D.}~\bibnamefont {Goldhaber-Gordon}},\ }\href@noop {} {\bibfield  {journal}
		{\bibinfo  {journal} {Nat. Phys.}\ }\textbf {\bibinfo {volume} {10}},\
		\bibinfo {pages} {145} (\bibinfo {year} {2014})}\BibitemShut {NoStop}%
	\bibitem [{\citenamefont {Mitchell}\ \emph {et~al.}(2021)\citenamefont
		{Mitchell}, \citenamefont {Liberman}, \citenamefont {Sela},\ and\
		\citenamefont {Affleck}}]{mitchell2021so}%
	\BibitemOpen
	\bibfield  {author} {\bibinfo {author} {\bibfnamefont {A.~K.}\ \bibnamefont
			{Mitchell}}, \bibinfo {author} {\bibfnamefont {A.}~\bibnamefont {Liberman}},
		\bibinfo {author} {\bibfnamefont {E.}~\bibnamefont {Sela}}, \ and\ \bibinfo
		{author} {\bibfnamefont {I.}~\bibnamefont {Affleck}},\ }\href@noop {}
	{\bibfield  {journal} {\bibinfo  {journal} {Phys. Rev. Lett.}\ }\textbf
		{\bibinfo {volume} {126}},\ \bibinfo {pages} {147702} (\bibinfo {year}
		{2021})}\BibitemShut {NoStop}%
	\bibitem [{\citenamefont {Liberman}\ \emph {et~al.}(2021)\citenamefont
		{Liberman}, \citenamefont {Mitchell}, \citenamefont {Affleck},\ and\
		\citenamefont {Sela}}]{liberman2021so}%
	\BibitemOpen
	\bibfield  {author} {\bibinfo {author} {\bibfnamefont {A.}~\bibnamefont
			{Liberman}}, \bibinfo {author} {\bibfnamefont {A.~K.}\ \bibnamefont
			{Mitchell}}, \bibinfo {author} {\bibfnamefont {I.}~\bibnamefont {Affleck}}, \
		and\ \bibinfo {author} {\bibfnamefont {E.}~\bibnamefont {Sela}},\ }\href@noop
	{} {\bibfield  {journal} {\bibinfo  {journal} {Phys. Rev. B}\ }\textbf
		{\bibinfo {volume} {103}},\ \bibinfo {pages} {195131} (\bibinfo {year}
		{2021})}\BibitemShut {NoStop}%
\end{thebibliography}

\end{document}